\documentclass[copyright]{eptcs}
\providecommand{\event}{DCM 2014 Post-Proceedings} 

\title{\scalebox{0.96}[1]{A Simple Parallel Implementation of Interaction Nets in Haskell}}
\author{Wolfram Kahl
\institute{Department of Computing and Software, McMaster University,
Hamilton, Ontario, Canada}
\email{kahl@cas.mcmaster.ca}
}
\def\titlerunning{A Simple Parallel Implementation of Interaction Nets in Haskell}
\def\authorrunning{Wolfram Kahl}

\makeatletter
\@ifundefined{lhs2tex.lhs2tex.sty.read}%
  {\@namedef{lhs2tex.lhs2tex.sty.read}{}%
   \newcommand\SkipToFmtEnd{}%
   \newcommand\EndFmtInput{}%
   \long\def\SkipToFmtEnd#1\EndFmtInput{}%
  }\SkipToFmtEnd

\newcommand\ReadOnlyOnce[1]{\@ifundefined{#1}{\@namedef{#1}{}}\SkipToFmtEnd}
\usepackage{amstext}
\usepackage{amssymb}
\usepackage{stmaryrd}
\DeclareFontFamily{OT1}{cmtex}{}
\DeclareFontShape{OT1}{cmtex}{m}{n}
  {<5><6><7><8>cmtex8
   <9>cmtex9
   <10><10.95><12><14.4><17.28><20.74><24.88>cmtex10}{}
\DeclareFontShape{OT1}{cmtex}{m}{it}
  {<-> ssub * cmtt/m/it}{}

\DeclareFontShape{OT1}{cmtt}{bx}{n}
  {<5><6><7><8>cmtt8
   <9>cmbtt9
   <10><10.95><12><14.4><17.28><20.74><24.88>cmbtt10}{}
\DeclareFontShape{OT1}{cmtex}{bx}{n}
  {<-> ssub * cmtt/bx/n}{}

\newcommand{\anonymous}{\kern0.06em \vbox{\hrule\@width.5em}}
\newcommand{\plus}{\mathbin{+\!\!\!+}}

\newcommand{\sequ}{\mathbin{>\!\!\!>}}

\usepackage{polytable}

\@ifundefined{mathindent}%
  {\newdimen\mathindent\mathindent\leftmargini}%
  {}%

\def\resethooks{%
  \global\let\SaveRestoreHook\empty
  \global\let\ColumnHook\empty}
\newcommand*{\savecolumns}[1][default]%
  {\g@addto@macro\SaveRestoreHook{\savecolumns[#1]}}
\newcommand*{\restorecolumns}[1][default]%
  {\g@addto@macro\SaveRestoreHook{\restorecolumns[#1]}}
\newcommand*{\aligncolumn}[2]%
  {\g@addto@macro\ColumnHook{\column{#1}{#2}}}

\resethooks

\newcommand{\onelinecommentchars}{\quad-{}- }
\newcommand{\commentbeginchars}{\enskip\{-}
\newcommand{\commentendchars}{-\}\enskip}

\newcommand{\visiblecomments}{%
  \let\onelinecomment=\onelinecommentchars
  \let\commentbegin=\commentbeginchars
  \let\commentend=\commentendchars}

\newcommand{\invisiblecomments}{%
  \let\onelinecomment=\empty
  \let\commentbegin=\empty
  \let\commentend=\empty}

\visiblecomments

\newlength{\blanklineskip}
\newcommand{\hsindent}[1]{\quad}
\let\hspre\empty
\let\hspost\empty

\EndFmtInput
\makeatother
\ReadOnlyOnce{polycode.fmt}%
\makeatletter

\newcommand{\hsnewpar}[1]%
  {{\parskip=0pt\parindent=0pt\par\vskip #1\noindent}}

\newcommand{\hscodestyle}{}

\newcommand{\sethscode}[1]%
  {\expandafter\let\expandafter\hscode\csname #1\endcsname
   \expandafter\let\expandafter\endhscode\csname end#1\endcsname}

  {\par\noindent
   \advance\leftskip\mathindent
   \hscodestyle
   \let\\=\@normalcr
   \let\hspre\(\let\hspost\)%
   \pboxed}%
  {\endpboxed\)%
   \par\noindent
   \ignorespacesafterend}

  {\hsnewpar\abovedisplayskip
   \advance\leftskip\mathindent
   \hscodestyle
   \let\hspre\(\let\hspost\)%
   \pboxed}%
  {\endpboxed%
   \hsnewpar\belowdisplayskip
   \ignorespacesafterend}

  {\hsnewpar\abovedisplayskip
   \advance\leftskip\mathindent
   \hscodestyle
   \let\\=\@normalcr
   \(\pboxed}%
  {\endpboxed\)%
   \hsnewpar\belowdisplayskip
   \ignorespacesafterend}

\newcommand{\plainhs}{\sethscode{plainhscode}}

\plainhs

  {\hsnewpar\abovedisplayskip
   \advance\leftskip\mathindent
   \hscodestyle
   \let\\=\@normalcr
   \(\parray}%
  {\endparray\)%
   \hsnewpar\belowdisplayskip
   \ignorespacesafterend}

  {\parray}{\endparray}

  {\(\parray}{\endparray\)}

\def\codeframewidth{\arrayrulewidth}
\RequirePackage{calc}

  {\parskip=\abovedisplayskip\par\noindent
   \hscodestyle
   \arrayrulewidth=\codeframewidth
   \tabular{@{}|p{\linewidth-2\arraycolsep-2\arrayrulewidth-2pt}|@{}}%
   \hline\framedhslinecorrect\\{-1.5ex}%
   \let\endoflinesave=\\
   \let\\=\@normalcr
   \(\pboxed}%
  {\endpboxed\)%
   \framedhslinecorrect\endoflinesave{.5ex}\hline
   \endtabular
   \parskip=\belowdisplayskip\par\noindent
   \ignorespacesafterend}

\newcommand{\framedhslinecorrect}[2]%
  {#1[#2]}

  {\(\def\column##1##2{}%
   \let\>\undefined\let\<\undefined\let\\\undefined
   \newcommand\>[1][]{}\newcommand\<[1][]{}\newcommand\\[1][]{}%
   \def\fromto##1##2##3{##3}%
   }{\) }%

  {\let\orighscode=\hscode
   \let\origendhscode=\endhscode
   \def\endhscode{\def\hscode{\endgroup\def\@currenvir{hscode}\\}\begingroup}
   \orighscode\def\hscode{\endgroup\def\@currenvir{hscode}}}%
  {\origendhscode
   \global\let\hscode=\orighscode
   \global\let\endhscode=\origendhscode}%

\makeatother
\EndFmtInput
\def\VarId#1{\mathsf{#1}}

\def\TRIGRAM#1#2#3#4#5{{#1}\kern-#2ex{#3}\kern-#4ex{#5}}

\long\def\ignore#1{}
\usepackage{comment}
\specialcomment{ModuleHead}{\par\begingroup\tiny}{\endgroup}
\excludecomment{ModuleHead}

\def\sectionname{Sect.\null{}}
\usepackage{sectref}

\usepackage{graphicx}
\let\citep=\cite
\let\citet=\cite
\def\etal{\textsl{et al.\null{}}}

\def\Inets{{I}\textsf{nets}}
\def\bang{\mathbf{!}}

\def\best#1{\hbox{\vBorder\vbox{\hBorder{\bf #1}\hBorder}\vBorder}}
\def\best#1{\fbox{\kern-0.5em{\bf #1}\kern-0.5em}}
\def\defaultHeap{{\bf dft.}}

\usepackage{hyperref}
\begin{document}
\renewcommand{\hscodestyle}{\small}
\maketitle

\begin{abstract}
Due to their ``inherent parallelism'',
interaction nets have since their introduction 
been considered as an attractive implementation mechanism
for functional programming.
We show that a simple highly-concurrent implementation
in Haskell can achieve promising speed-ups on multiple cores.
\end{abstract}

\section{Introduction}

The \emph{interaction nets} introduced by Lafont \cite{Lafont-1990}
can be considered as a variant of term graphs,
and therewith as a kind of graphs used as representation of terms.
Interaction nets are equipped with an ``inherently parallel''
local and confluent reduction mechanism
that makes them an, at least conceptually, attractive target
for (functional) programming language implementation.
However, to date there have been only limited experiments with
parallel implementations of interaction nets,
and no easily-usable parallel implementation is publicly available.
In addition, the nature of the parallelism of interaction net
reduction is in general rather fine-grained,
so that the question of distribution strategies arises naturally.

In this paper, we report on an experiment that bypasses the question
of distribution strategies, and instead investigates whether
a fine-grained threading mechanism with parallel execution on
shared-memory multi-core systems,
as provided by the run-time system of the Glasgow Haskell Compiler (GHC),
can already realise the potential of parallelisation offered by
interaction nets.
Our implementation is publicly available
(at \url{http://www.cas.mcmaster.ca/~kahl/Haskell/HINet/})
and accepts a slightly restricted version of the \Inets{} file format,
enabling further experiments also by other interaction net researchers.
In the benchmarking section,
we provide a lot of data, and also discuss the potential
pitfalls of benchmarking Haskell programs
with large heap requirements,
in order to aid potential users of our system to avoid these pitfalls.

\subsection{From Term Graphs via Jungles and Code Graphs to Interaction Nets}

We now give an introduction to interaction nets
that puts them into the context of different term graph
representations.
We do this for two reasons:
First, to make interaction nets more accessible for readers interested in
functional programming language implementation, who may already be familiar
with graph reduction, but might find the principal-port orientation of
most of the interaction net literature rather obscure,
and second, to give a clear understanding of polarities, which have
almost disappeared from the interaction net literature.

Conventional term graphs (see e.g. \citep{Kennaway-Klop-Sleep-deVries-1993a})
are node-labelled directed graphs,
where each node has a sequence of outgoing edge
the length of which is determined (or sometimes part of) the label.
Node labels of these term graphs correspond to function symbols in
terms;
variables do not need labels: Different variable nodes
(labelled ``\textsf{\itshape V}'' below)
represent different variables.

The ``jungle'' approach of Hoffmann and Plump \citep{Hoffmann-Plump-1991}
moves the function symbols into hyperedges,
with a sequence of ``argument tentacles'' (or ``input tentacles'')
extending to argument nodes,
and (normally) exactly one ``result tentacle'' (or ``output tentacle'')
extending to the hyperedge's result node (or output node).
In both approaches, there is no restriction on the
number of edges (resp.\null{} input tentacles) incoming into each node;
multiple incoming edges implement \emph{sharing}
(and zero incoming edges into a non-root node
implement (uncollected) ``garbage'',
where in term graph and jungle rewriting,
garbage collection is typically implicit).

\kern-1ex
\strut\hfill
\includegraphics[scale=0.7,viewport=97 606 255 739]{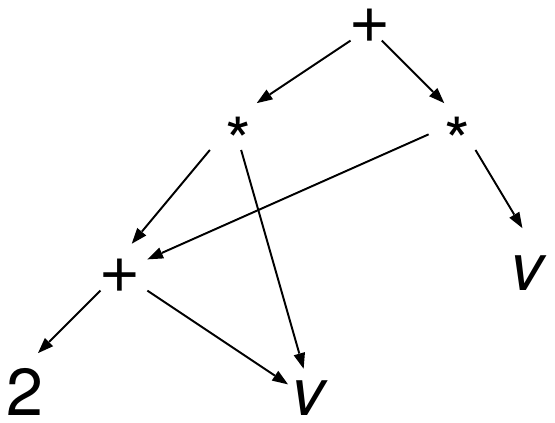}
\hfill
\includegraphics[scale=0.7,viewport=27 610 174 735]{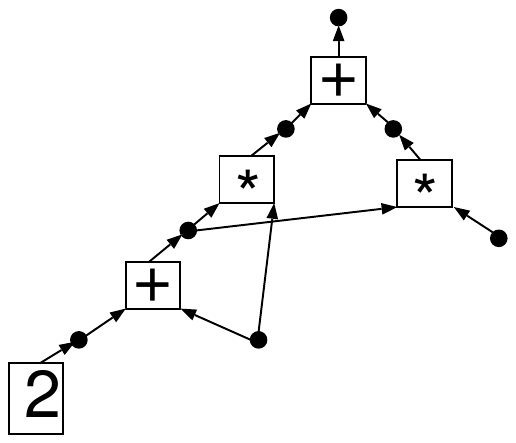}
\hfill
\includegraphics[scale=0.7,viewport=54 608 178 754]{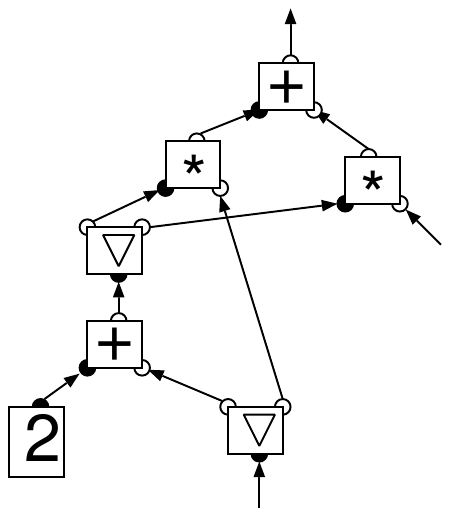}
\hfill\strut

\strut\hfill
{Term Graph}
\hfill
{Jungle}
\hfill
{Interaction Net}
\hfill\strut

\kern1ex
\noindent
The drawing above shows a conventional term graph,
a jungle, and an interaction net
each representing the term $(2 + x) * x + (2 + x) * y$
with the same degree of sharing.
In all three drawings, the sequence of the outgoing or incoming edges,
respectively tentacles, or ports, of each node or hyperedge is part of the structure,
but is, as customary, not made more explicit.

\medbreak
Interaction nets are different from jungles in several ways.
First of all, a different terminology is used:
  Instead of ``hyperedge'',
  the terms ``node'' or ``agent'' are used,
  the nodes of jungles turn into ``connections''
  and the tentacle labels and directions turn into ``ports''.
In interaction nets, connections must be incident with exactly one or two ports;
  those incident with only one port make up the interface of the net.
  Because of this, sharing and garbage must be made explicit
  via duplicator (``$\nabla$'') and terminator (``$\bang$'') nodes.
Each interaction net node label determines one \emph{principal port} for its
  nodes. We draw principal ports as filled-in circles attached to
  the rectangular nodes, while auxiliary ports are hollow.
Interaction net rules only replace pairs of nodes connected via their
  principal ports.

The directions of edges in termgraphs,
  and of tentacles in jungles,
  are motivated by denotational semantics;
  the corresponding 
  directions
  of connections in interaction nets
  were introduced under the name \emph{polarities} by Lafont
  \cite{Lafont-1990},
  but are omitted in a large part of the interaction net literature,
  where interaction nets are drawn with undirected connections.
  Instead, the operationally motivated direction of nodes (``actors'')
  from auxiliary ports to the principal port is typically emphasised.
  We follow Lafont \citet{Lafont-1990}
  to distinguish output ports (with positive \emph{polarity})
  and input ports (negative polarity),
  and draw connections as directed arrows from output to input ports.
  Note that besides Lafont \citet{Lafont-1990},
  most of the interaction net literature does \emph{not} draw nets in a way
  that easily corresponds to a jungle reading.

Whereas jungle hyperedges have only one output tentacle,
the duplicator ($\nabla$) nodes of the interaction net above
have two output ports --- a feature that also occurs in the
\emph{code graphs} of \cite{Kahl-Anand-Carette-2005,Anand-Kahl-2009a}.
We illustrate this with a second example;
the
term $(2 / x) * y + (2 \% x) * y$ represented with sharing as a term
graph has two variable nodes corresponding to $x$ and $y$;
represented as jungle these turn into two input nodes.

\kern-1.0ex
\strut\hfill
\includegraphics[scale=0.7,viewport=106 606 237 727]{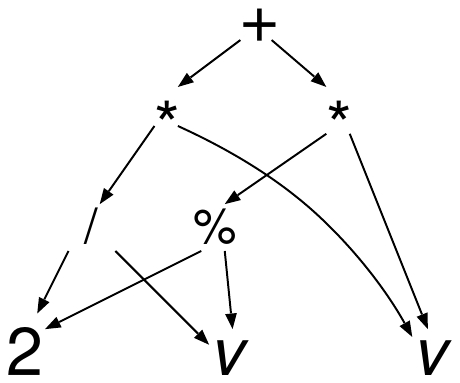}
\hfill
\includegraphics[scale=0.7,viewport=27 609 160 735]{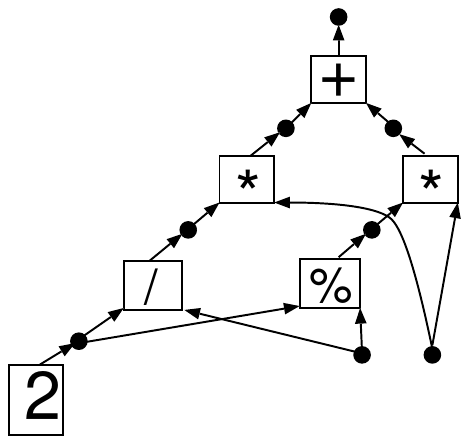}
\hfill\strut

\strut\hfill
{Term Graph}
\hfill
{Jungle}
\hfill\strut

\kern1.4ex
\noindent
In code graphs, the sequence of these input nodes is explicitly
visualised via triangular tags with arrows towards the input nodes;
code graphs also have a sequence of output nodes
visualised via triangular tags with arrows from the output nodes.
Code graph hyperedges also have as interface a sequence of input nodes
(as in jungles)
and a sequence of output nodes, which in contrast to jungles is not
constrained to contain exactly one element.
For the sake of an example, we can therefore use a two-output operation ``$\mathsf{divMod}$''
to obtain a code graph that uses a single operation to produce the
same result as the two separate operations $/$ and $\%$
in the term and jungle above.
(The sequences of input and output nodes of hyperedges are still indicated implicitly
via the graphical arrangement.)

\kern1.0ex
\strut\hfill
\includegraphics[scale=0.7,viewport=66 609 173 757]{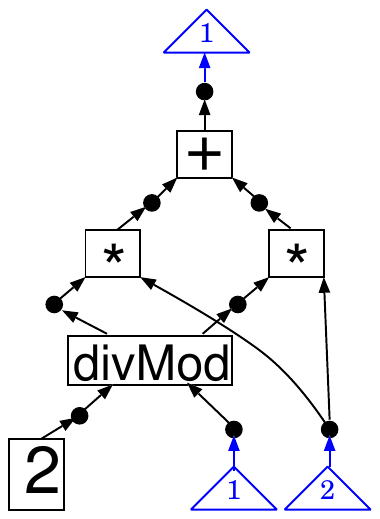}
\hfill
\includegraphics[scale=0.7,viewport=66 608 168 755]{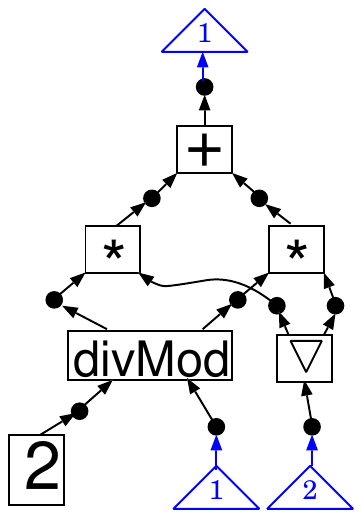}
\hfill
\raise1.7ex\hbox{\includegraphics[scale=0.7,viewport=60 643 187 763]{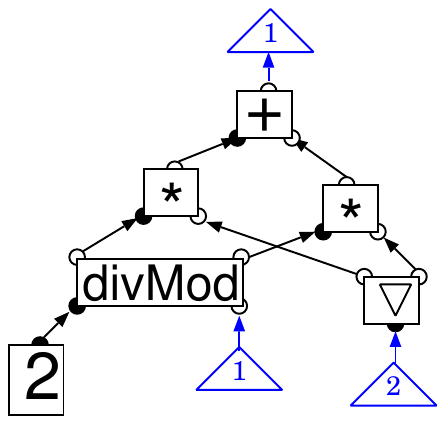}}
\hfill\strut

\strut\hfill
{Code Graph}
\hfill
{Code Graph with Duplicator}
\hfill
{Interaction Net}
\hfill
\strut

\kern1.0ex
\noindent
Since code graphs allow multi-output nodes,
duplicators (``$\nabla$'') do not need to be given any special status,
and interaction net languages can be understood
as code graph languages without node-based sharing (and without
``garbage''), which allows us to replace the code graph nodes
with their single incoming and outgoing tentacles with
simple connections.
Input and output nodes of code graphs
turn into input and output ports of interaction nets ---
these are the ports of negative, respectively positive polarity
that have no connection attached to them.
As for code graphs, we will assume the input and output ports to be
organised into two sequences, and tag them using the same triangles.

\subsection{Interaction Net Rules and Reduction}

Application of rules is defined as subnet replacement,
where the input and output ports of the rule sides
may map to arbitrary ports in the application net.
Due to the constraints on the left-hand sides of rules,
the resulting reduction has no critical pairs;
it is therefore confluent
and has a deterministic normalisation relation.
Since left-hand sides match only to subnets induced by
two nodes connected via their principal ports,
reduction exhibits extreme locality,
and is frequently considered as ``inherently parallel''.

Below, we show rules for addition and multiplication of natural
numbers
built up from the constructors for zero (``$\mathsf{0}$'') and
successor (``$\mathsf{S}$'').
The first multiplication rule, ``\ensuremath{\VarId{mult}\;\mathrm{0}\;\VarId{n}\mathrel{=}\mathrm{0}}'', turns \ensuremath{\VarId{n}} into
``garbage'' by attaching a terminator (``$\bang$'') node;
the second multiplication rule ``duplicates'' \ensuremath{\VarId{n}} for use both by the
addition and by the recursive call.

\kern-0.3ex
\strut\quad\begin{minipage}[b]{0.4\columnwidth}
\begin{hscode}\SaveRestoreHook
\column{B}{@{}>{\hspre}l<{\hspost}@{}}%
\column{E}{@{}>{\hspre}l<{\hspost}@{}}%
\>[B]{}\VarId{add}\;\mathrm{0}\;\VarId{n}\mathrel{=}\VarId{n}{}\<[E]%
\\
\>[B]{}\VarId{add}\;(\VarId{S}\;\VarId{m})\;\VarId{n}\mathrel{=}\VarId{S}\;(\VarId{add}\;\VarId{m}\;\VarId{n}){}\<[E]%
\ColumnHook
\end{hscode}\resethooks
\end{minipage}
\hfill
\begin{minipage}[b]{0.46\columnwidth}
\begin{hscode}\SaveRestoreHook
\column{B}{@{}>{\hspre}l<{\hspost}@{}}%
\column{E}{@{}>{\hspre}l<{\hspost}@{}}%
\>[B]{}\VarId{mult}\;\mathrm{0}\;\VarId{n}\mathrel{=}\mathrm{0}{}\<[E]%
\\
\>[B]{}\VarId{mult}\;(\VarId{S}\;\VarId{m})\;\VarId{n}\mathrel{=}\VarId{add}\;\VarId{n}\;(\VarId{mult}\;\VarId{m}\;\VarId{n}){}\<[E]%
\ColumnHook
\end{hscode}\resethooks
\end{minipage}

\kern-1ex
\noindent
\strut
\includegraphics[scale=0.7,viewport=99 645 229 763]{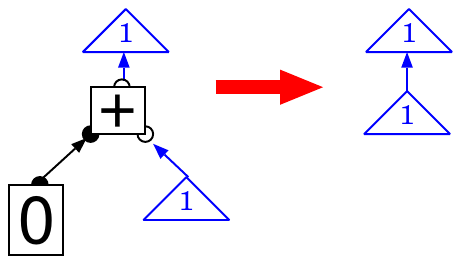}
\hfill\vrule height17ex width0.3pt depth0.2ex\hfill
\includegraphics[scale=0.7,viewport=95 645 246 763]{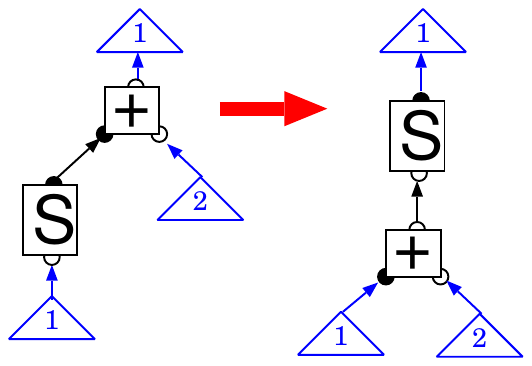}
\hfill\smash{\vrule height24ex width0.3pt depth0.2ex}\hfill
\includegraphics[scale=0.7,viewport=99 645 258 763]{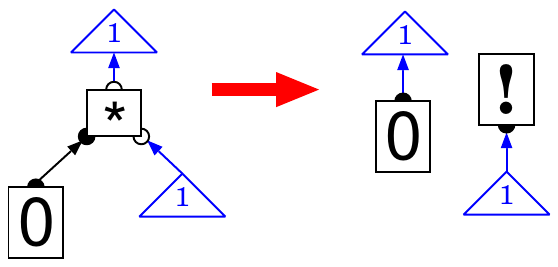}
\hfill\vrule height17ex width0.3pt depth0.2ex\hfill
\includegraphics[scale=0.7,viewport=95 645 265 763]{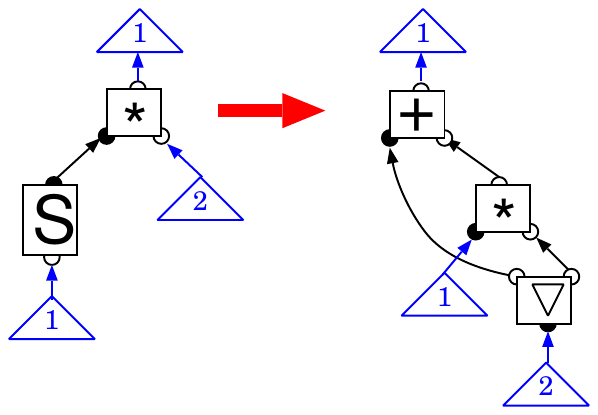}
\strut

\kern1ex
\noindent
The reader may notice that the multiplication rules provided above
always perform a ``superfluous'' last addition to zero
if the first factor is non-zero.
One might consider the following starting point instead:

\kern1ex
\strut\kern-3em
\begin{minipage}[b]{0.34\columnwidth}
\begin{hscode}\SaveRestoreHook
\column{B}{@{}>{\hspre}l<{\hspost}@{}}%
\column{E}{@{}>{\hspre}l<{\hspost}@{}}%
\>[B]{}\VarId{mult}\;\mathrm{0}\;\VarId{n}\mathrel{=}\mathrm{0}{}\<[E]%
\\
\>[B]{}\VarId{mult}\;(\VarId{S}\;\mathrm{0})\;\VarId{n}\mathrel{=}\VarId{n}{}\<[E]%
\\
\>[B]{}\VarId{mult}\;(\VarId{S}\;\VarId{m})\;\VarId{n}\mathrel{=}\VarId{add}\;\VarId{n}\;(\VarId{mult}\;\VarId{m}\;\VarId{n}){}\<[E]%
\ColumnHook
\end{hscode}\resethooks
\end{minipage}
\hfill
\includegraphics[scale=0.6,viewport=99 645 258 763]{Drawings/mult-zero_inet}
\hfill\vrule height16ex width0.3pt depth0.2ex\hfill
\includegraphics[scale=0.6,viewport=98 638 229 764]{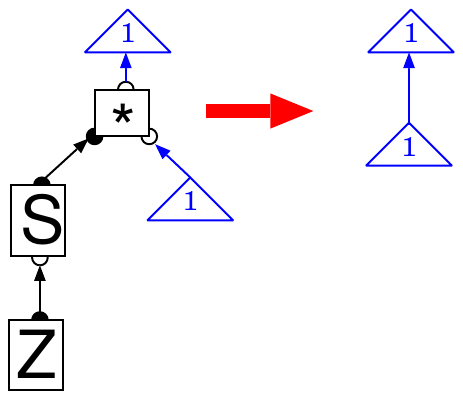}
\hfill\vrule height16ex width0.3pt depth0.2ex\hfill
\includegraphics[scale=0.6,viewport=94 628 265 764]{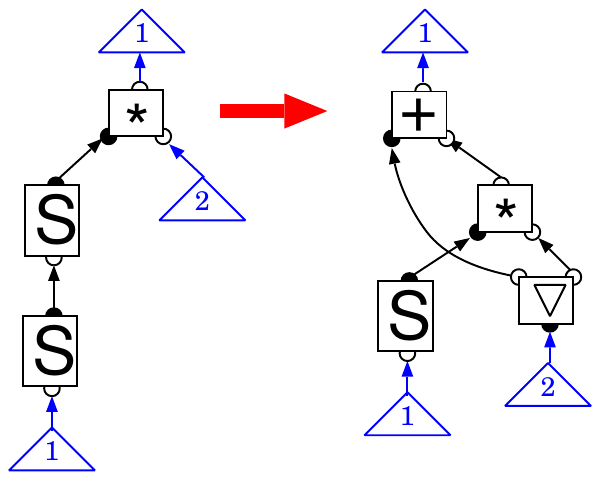}
\strut

\noindent
However, the ``deep pattern matching'' here cannot be implemented
directly by conventional interaction net reduction;
the rules drawn above to the right
are however allowed in the extension proposed by Hassan \etal{}
\cite{Hassan-Jiresch-SatoShinya-2010}
which translates them into conventional interaction net rules by adding
an auxiliary function:

\strut\kern-3em
\begin{minipage}[b]{0.23\columnwidth}
\begin{hscode}\SaveRestoreHook
\column{B}{@{}>{\hspre}l<{\hspost}@{}}%
\column{3}{@{}>{\hspre}l<{\hspost}@{}}%
\column{E}{@{}>{\hspre}l<{\hspost}@{}}%
\>[B]{}\VarId{mult}\;\mathrm{0}\;\VarId{n}\mathrel{=}\mathrm{0}{}\<[E]%
\\
\>[B]{}\VarId{mult}\;(\VarId{S}\;\VarId{m})\;\VarId{n}{}\<[E]%
\\
\>[B]{}\hsindent{3}{}\<[3]%
\>[3]{}\mathrel{=}\VarId{multAux}\;\VarId{m}\;\VarId{n}{}\<[E]%
\\
\>[B]{}\VarId{multAux}\;\mathrm{0}\;\VarId{n}\mathrel{=}\VarId{n}{}\<[E]%
\\
\>[B]{}\VarId{multAux}\;(\VarId{S}\;\VarId{m})\;\VarId{n}{}\<[E]%
\\
\>[B]{}\hsindent{3}{}\<[3]%
\>[3]{}\mathrel{=}\VarId{add}\;\VarId{n}\;(\VarId{multAux}\;\VarId{m}\;\VarId{n}){}\<[E]%
\ColumnHook
\end{hscode}\resethooks

\kern-5ex
\end{minipage}
\hfill
\includegraphics[scale=0.6,viewport=76 665 286 764]{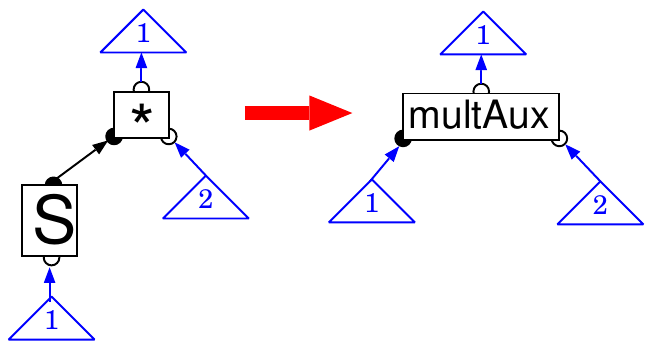}
\hfill\vrule height13ex width0.3pt depth0.2ex\hfill
\includegraphics[scale=0.6,viewport=85 665 249 764]{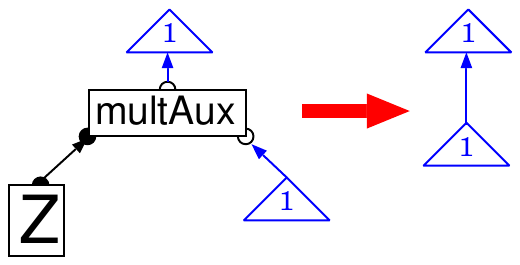}
\hfill\vrule height13ex width0.3pt depth0.2ex\hfill
\includegraphics[scale=0.6,viewport=62 647 277 764]{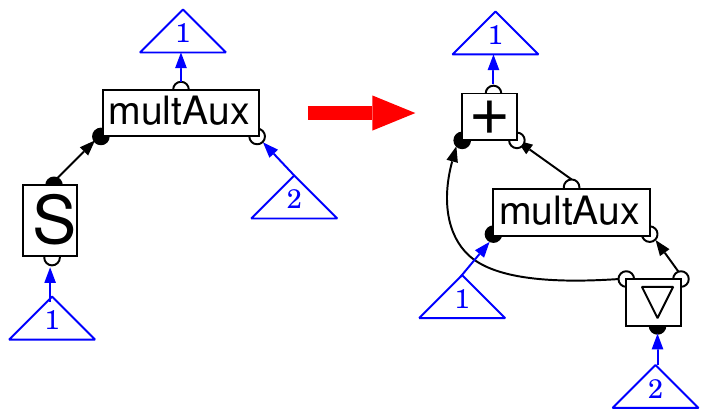}

\kern3ex
\noindent
Such encoding issues are not relevant to the current study,
which considers
interaction nets as an execution model,
rather than as a programming language.
Compilation to interaction net rules is a separate topic,
and has been studied for example
by H.~Cirstea and others
\cite{CirsteaH-Faure-Fernandez-Mackie-Sinot-2007}
using the $\rho$-calculus as intermediate language.

\subsection{Related Work}

Pedicini and Quaglia \cite{Pedicini-Quaglia-2007} describe PELCR, a
distributed parallel environment for optimal $\lambda$-calculus
reduction, which uses a specialised fixed interaction net language
and implements sophisticated distribution strategies.
(I found no trace of this being or having been publicly available.)
Besides such specialised systems,
we are aware of only a small number of parallel implementations of
interaction nets,
in particular
\cite{Banach-Papadopoulos-1997,Pinto-2001RTA,Jiresch-2014}.
Of all these, only the last seems to be (still) available;
it is an experimental GPU implementation that requires
new rules to be implemented manually in C/CUDA at a very low level.

A general interaction net implementation that is still available
is part of the \Inets{} project of Mackie \etal{}
\cite{Hassan-Mackie-Sato-2009,INets}.
This it is a compiler
for the interaction net definition language \Inets{},
which is considered as a programming language;
the compiler is implemented in Java, and compiles via C to non-parallel
executables.
While \Inets{} implements nets as pointer structures,
the (apparently unavailable) successor system ``Light''
\cite{Hassan-Mackie-Sato-2010}, as well as
the systems of Pinto \cite{Pinto-2001RTA} and Jiresch \cite{Jiresch-2014}
are based on a term representation of interaction nets
(based on the fact already pointed out by Lafont \cite{Lafont-1990}
that ``well-behaved'' fully reduced nets always can be represented
via pairs of terms with common variables and further constraints).
Lippi's implementation called ``$\mathsf{in}^2$'' \cite{Lippi-2002}
was apparently close in spirit, but not directly based on terms.

Other available implementations
are geared more towards graphical interaction directly with interaction
nets (and also don't support parallel execution),
including de Falco's ``Interaction Nets Laboratory''
\cite{deFalco-2006},
the ``interaction net IDE'' INblobs of Almeida \etal{}
\cite{Almeida-Pinto-Vilaca-2008},
and the graph rewriting system IDE ``PORGY''
\cite{Andrei-Fernandez-Kirchner-Melancon-Namet-Pinaud-2011}
which can also be used for interaction nets.
By emphasising visualisation of net transformations,
these tools by design cannot target efficient parallel implementation.

\subsection{Contribution and Overview}

We present a design for highly concurrent interaction net
implementations that is at the same time surprisingly simple
and very close to the graph understanding of the interaction net
definition.
The parallel implementation of concurrency in the Glasgow Haskell
Compiler (GHC) is a good fit for this kind of design;
our implementation obtains satisfactory speed-ups even for simple
examples.


While most current non-graphical implementations of interaction nets
are based on a term-based calculus,
we explain our more direct approach in \sectref{ImplementationDesign}.
The actual (literate) Haskell source
code of the kernel of our implementation is then presented in
\sectref{Implementation} --- the full source code
is available on-line at
\url{http://www.cas.mcmaster.ca/~kahl/Haskell/HINet/}.
In \sectref{RunInets} we summarise our implementation of a
language
similar to that of \Inets{} \cite{INets}.
Measurements and relevant observations are in
\sectref{Benchmarks}.

\section{Implementation Design}\sectlabel{ImplementationDesign}

\noindent
Our implementation essentially follows the main ideas of
Banach and Papadopoulos \cite{Banach-Papadopoulos-1997}:
\begin{itemize}
\item
Two-way connections, which easily introduce opportunities for deadlock
  and race conditions, can be avoided by using polarities
  to direct the connections between ports
  (which, in a large part of the literature, are treated as undirected, and implemented
  as two-way connections).
\item
 These \emph{directed} connections hold mutable state.
\item
  The connection with the principal port of a constructor
  does not need to be known to the constructor node
  if the connection state refers to the node.
\end{itemize}
The following main decisions then determine most of our implementation
details:

\kern-0.5ex
\begin{itemize}\itemsep0pt
\item Connections (drawn below as thick circles)
  are initially ``empty'', and each node
  has references to the connections attached to its auxiliary ports.
\item Attaching the principal port of a constructor to a connection
  deposits a reference to the constructor node in the connection
  (which is then ``full'').
  (This reference is drawn below with a thick arrow with a bullet tail.)
\item Attaching the principal port of a function to a connection
  starts a concurrent thread that waits for a constructor reference
  in that connection, and if/when it finds one,
  starts the corresponding rule application.
  (This is drawn below with an even thicker arrow ending inside the connection.)
\end{itemize}

\kern-0.4ex
\noindent
The following shows a net fragment first in the same style as the
previous example, and to the right with implementation details added.

\kern1ex
\strut\hfill
\raise3ex\hbox{\includegraphics[scale=0.8,viewport=54 608 175 712]{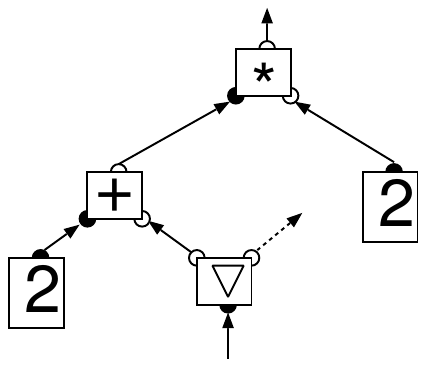}}
\hfill
\includegraphics[scale=0.7,viewport=24 595 238 741]{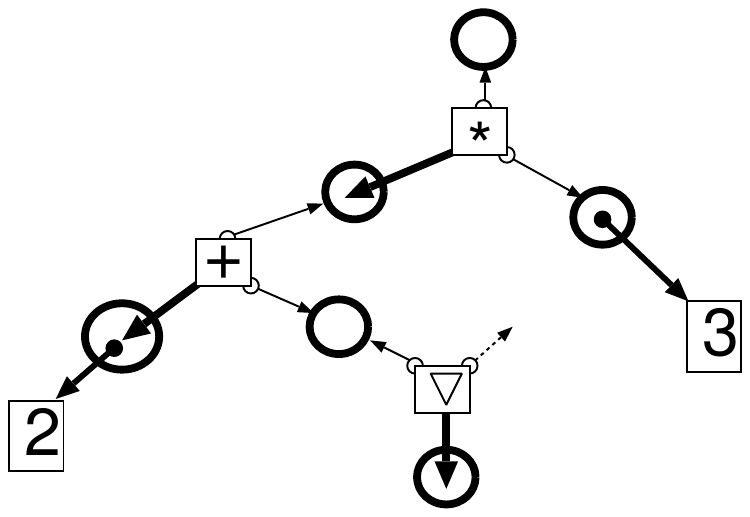}
\hfill\strut

\section{Implementation in Concurrent Haskell}\sectlabel{Implementation}

\noindent
We implement connections using the Concurrent Haskell synchronisation
primitive \ensuremath{\VarId{MVar}}, which can be created empty;
\ensuremath{\VarId{putMVar}} waits for empty state to fill, and \ensuremath{\VarId{takeMVar}}
waits for full state to empty
\cite{PeytonJones-Gordon-Finne-1996}.
The GHC version of Concurrent Haskell has an extremely
light-weight thread implementation that makes it feasible
to create millions of threads; we therefore directly
create new threads for functions as mentioned above,
and even smaller threads for short-circuiting two interface
ports that are directly connected by rule applications:
These threads only wait for a constructor on the originally negative
port of the LHS, and copy it to the positive side.

The run-time implementation of nets, based on \ensuremath{\VarId{MVar}}s,
is introduced in \sectref{INet.Polar.INet}.
For the static representation of rules,
our implementation uses a non mutable datatype \ensuremath{\VarId{NetDescription}}
to represent right-hand sides (RHSs) of reduction rules;
these are introduced in \sectref{INet.Description}.
At run-time, these \ensuremath{\VarId{NetDescription}}s are instantiated into new parts of the mutable
run-time net, as fully defined in \sectref{INet.Polar.Reduce}
following the principles outlined in \sectref{ImplementationDesign}.

{\def\section#1{\subsection{#1}}
\section{Polarity}

\begin{ModuleHead}
\begin{hscode}\SaveRestoreHook
\column{B}{@{}>{\hspre}l<{\hspost}@{}}%
\column{E}{@{}>{\hspre}l<{\hspost}@{}}%
\>[B]{}\mathbf{module}\;\VarId{\VarId{INet}.Polarity}\;\mathbf{where}{}\<[E]%
\ColumnHook
\end{hscode}\resethooks
\end{ModuleHead}

Lafont \cite{Lafont-1990} and Banach and Papadopoulos \cite{Banach-Papadopoulos-1997}
use typed connections in their interaction nets,
where the two ports incident in a connection have the same type,
but different polarity.
Since we design our interaction net implementation as a run-time system,
types are currently not important,
and will be assumed to have been taken care of before net generation.
Polarity, however, drives several run-time decisions;
for the sake of readability, we define a special-purpose data-type for it
(and let Haskell's ``\ensuremath{\mathbf{deriving}}'' mechanism provide us
with the default implementation of equality and ordering tests,
and of conversion to strings):

\begin{hscode}\SaveRestoreHook
\column{B}{@{}>{\hspre}l<{\hspost}@{}}%
\column{3}{@{}>{\hspre}l<{\hspost}@{}}%
\column{E}{@{}>{\hspre}l<{\hspost}@{}}%
\>[B]{}\mathbf{data}\;\VarId{Polarity}\mathrel{=}\VarId{Neg}\mid \VarId{Pos}{}\<[E]%
\\
\>[B]{}\hsindent{3}{}\<[3]%
\>[3]{}\mathbf{deriving}\;(\VarId{Eq},\VarId{Ord},\VarId{Show}){}\<[E]%
\\[\blanklineskip]%
\>[B]{}\VarId{opposite}\mathbin{::}\VarId{Polarity}\to \VarId{Polarity}{}\<[E]%
\\
\>[B]{}\VarId{opposite}\;\VarId{Neg}\mathrel{=}\VarId{Pos}{}\<[E]%
\\
\>[B]{}\VarId{opposite}\;\VarId{Pos}\mathrel{=}\VarId{Neg}{}\<[E]%
\ColumnHook
\end{hscode}\resethooks

\noindent
We will follow Lafont's convention of letting ``constructors''
have positive polarity, and ``functions'' negative polarity.

\section{Mutable Net Representation}\sectlabel{INet.Polar.INet}

\begin{ModuleHead}
\begin{hscode}\SaveRestoreHook
\column{B}{@{}>{\hspre}l<{\hspost}@{}}%
\column{E}{@{}>{\hspre}l<{\hspost}@{}}%
\>[B]{}\mathbf{module}\;\VarId{\VarId{INet}.\VarId{Polar}.INet}\;\mathbf{where}{}\<[E]%
\\[\blanklineskip]%
\>[B]{}\mathbf{import}\;\VarId{\VarId{Control}.\VarId{Concurrent}.MVar}{}\<[E]%
\\
\>[B]{}\mathbf{import}\;\VarId{\VarId{INet}.\VarId{Utils}.Vector}\;(\VarId{Vector}){}\<[E]%
\\
\>[B]{}\mathbf{import}\;\VarId{\VarId{INet}.Polarity}\;(\VarId{Polarity},\VarId{opposite}){}\<[E]%
\ColumnHook
\end{hscode}\resethooks
\end{ModuleHead}

A connection between two ports is implemented as a single \ensuremath{\VarId{MVar}}
that is either empty, or
contains the constructor node for which the connection is at the principal port.
(To allow different node label types to be used, we use the type variable \ensuremath{\VarId{nLab}} throughout.)
\begin{hscode}\SaveRestoreHook
\column{B}{@{}>{\hspre}l<{\hspost}@{}}%
\column{E}{@{}>{\hspre}l<{\hspost}@{}}%
\>[B]{}\mathbf{type}\;\VarId{Conn}\;\VarId{nLab}\mathrel{=}\VarId{MVar}\;(\VarId{Node}\;\VarId{nLab}){}\<[E]%
\ColumnHook
\end{hscode}\resethooks
For an auxiliary port of a node,
besides its connection
we also record the port's polarity to make it available efficiently at run-time.
(In Haskell, data constructors for simple record types
 habitually are given the same name as the type constructor;
 the fields \ensuremath{\VarId{pol}} and \ensuremath{\VarId{conn}} here are declared strict using ``\ensuremath{\mathbin{!}}'', and
 the ``\ensuremath{\VarId{UNPACK}}'' pragma declares an ``unpacking'' optimisation
 as desired to the compiler.)
%
\begin{hscode}\SaveRestoreHook
\column{B}{@{}>{\hspre}l<{\hspost}@{}}%
\column{3}{@{}>{\hspre}l<{\hspost}@{}}%
\column{11}{@{}>{\hspre}l<{\hspost}@{}}%
\column{30}{@{}>{\hspre}l<{\hspost}@{}}%
\column{E}{@{}>{\hspre}l<{\hspost}@{}}%
\>[B]{}\mathbf{data}\;\VarId{Port}\;\VarId{nLab}\mathrel{=}\VarId{Port}{}\<[E]%
\\
\>[B]{}\hsindent{3}{}\<[3]%
\>[3]{}\{\mskip1.5mu \VarId{pol}{}\<[11]%
\>[11]{}\mathbin{::}{}\<[30]%
\>[30]{}\mathbin{!}\VarId{Polarity}{}\<[E]%
\\
\>[B]{}\hsindent{3}{}\<[3]%
\>[3]{},\VarId{conn}{}\<[11]%
\>[11]{}\mathbin{::}\mbox{\enskip\{-\# UNPACK  \#-\}\enskip}{}\<[30]%
\>[30]{}\mathbin{!}(\VarId{Conn}\;\VarId{nLab}){}\<[E]%
\\
\>[B]{}\hsindent{3}{}\<[3]%
\>[3]{}\mskip1.5mu\}{}\<[E]%
\ColumnHook
\end{hscode}\resethooks
We introduce the type synonym \ensuremath{\VarId{Ports}} to abbreviate the type of port arrays.
\begin{hscode}\SaveRestoreHook
\column{B}{@{}>{\hspre}l<{\hspost}@{}}%
\column{13}{@{}>{\hspre}l<{\hspost}@{}}%
\column{E}{@{}>{\hspre}l<{\hspost}@{}}%
\>[B]{}\mathbf{type}\;\VarId{Ports}\;{}\<[13]%
\>[13]{}\VarId{nLab}\mathrel{=}\VarId{Vector}\;(\VarId{Port}\;\VarId{nLab}){}\<[E]%
\ColumnHook
\end{hscode}\resethooks
Given a port \ensuremath{\VarId{p}}, the port at the other end of its connection
is obtained as \ensuremath{\VarId{opPort}\;\VarId{p}} by flipping the polarity:
%
\begin{hscode}\SaveRestoreHook
\column{B}{@{}>{\hspre}l<{\hspost}@{}}%
\column{E}{@{}>{\hspre}l<{\hspost}@{}}%
\>[B]{}\VarId{opPort}\mathbin{::}\VarId{Port}\;\VarId{nLab}\to \VarId{Port}\;\VarId{nLab}{}\<[E]%
\\
\>[B]{}\VarId{opPort}\;\VarId{p}\mathrel{=}\VarId{p}\;\{\mskip1.5mu \VarId{pol}\mathrel{=}\VarId{opposite}\mathbin{\$}\VarId{pol}\;\VarId{p}\mskip1.5mu\}{}\<[E]%
\ColumnHook
\end{hscode}\resethooks
A node contains a label,
and the array of its \emph{non-principal} ports.
We do not include the principal port in \ensuremath{\VarId{ports}} since
\begin{itemize}
\item
the principal port of a constructor is connected to the \ensuremath{\VarId{MVar}}
  pointing back to the constructor,
  and
\item
 the principal port of a function is connected to the \ensuremath{\VarId{MVar}}
  the function's thread is waiting on.
\end{itemize}
\begin{hscode}\SaveRestoreHook
\column{B}{@{}>{\hspre}l<{\hspost}@{}}%
\column{3}{@{}>{\hspre}l<{\hspost}@{}}%
\column{E}{@{}>{\hspre}l<{\hspost}@{}}%
\>[B]{}\mathbf{data}\;\VarId{Node}\;\VarId{nLab}\mathrel{=}\VarId{Node}{}\<[E]%
\\
\>[B]{}\hsindent{3}{}\<[3]%
\>[3]{}\{\mskip1.5mu \VarId{label}\mathbin{::}\VarId{nLab}{}\<[E]%
\\
\>[B]{}\hsindent{3}{}\<[3]%
\>[3]{},\VarId{ports}\mathbin{::}\VarId{Ports}\;\VarId{nLab}{}\<[E]%
\\
\>[B]{}\hsindent{3}{}\<[3]%
\>[3]{}\mskip1.5mu\}{}\<[E]%
\ColumnHook
\end{hscode}\resethooks


\section{Net Descriptions}\sectlabel{INet.Description}

\begin{ModuleHead}
\begin{hscode}\SaveRestoreHook
\column{B}{@{}>{\hspre}l<{\hspost}@{}}%
\column{19}{@{}>{\hspre}l<{\hspost}@{}}%
\column{27}{@{}>{\hspre}l<{\hspost}@{}}%
\column{E}{@{}>{\hspre}l<{\hspost}@{}}%
\>[B]{}\mathbf{module}\;\VarId{\VarId{INet}.Description}\;\mathbf{where}{}\<[E]%
\\[\blanklineskip]%
\>[B]{}\mathbf{import}\;\VarId{\VarId{INet}.Polarity}{}\<[E]%
\\
\>[B]{}\mathbf{import}\;\VarId{\VarId{INet}.\VarId{Utils}.Vector}\;{}\<[27]%
\>[27]{}(\VarId{Vector},(\mathbin{!?}),\VarId{atErr},\VarId{bounds},(\mathbin{!})){}\<[E]%
\\
\>[B]{}\mathbf{import}\;\VarId{qualified}\;\VarId{\VarId{INet}.\VarId{Utils}.Vector}\;\VarId{as}\;\VarId{V}{}\<[E]%
\\[\blanklineskip]%
\>[B]{}\mathbf{import}\;\VarId{\VarId{Data}.List}{}\<[E]%
\\[\blanklineskip]%
\>[B]{}\mathbf{import}\;\VarId{\VarId{Data}.\VarId{Map}.Strict}\;(\VarId{Map}){}\<[E]%
\\
\>[B]{}\mathbf{import}\;\VarId{qualified}\;{}\<[19]%
\>[19]{}\VarId{\VarId{Data}.\VarId{Map}.Strict}\;\VarId{as}\;\VarId{Map}{}\<[E]%
\ColumnHook
\end{hscode}\resethooks
\end{ModuleHead}

Whereas in \sectref{INet.Polar.INet}, we introduced types for nets
considered as run-time states,
here we introduce \emph{net description}
for static representation of, in particular, rule right-hand sides.

The following types are dictated by our current choice of
array implementation (\ensuremath{\VarId{\VarId{Data}.Vector}} from the \ensuremath{\VarId{vector}} package, for efficiency),
but aliased for readability:
\begin{hscode}\SaveRestoreHook
\column{B}{@{}>{\hspre}l<{\hspost}@{}}%
\column{10}{@{}>{\hspre}l<{\hspost}@{}}%
\column{E}{@{}>{\hspre}l<{\hspost}@{}}%
\>[B]{}\mathbf{type}\;\VarId{PI}{}\<[10]%
\>[10]{}\mathrel{=}\VarId{Int}\mbox{\onelinecomment  ``port index''}{}\<[E]%
\\
\>[B]{}\mathbf{type}\;\VarId{NI}{}\<[10]%
\>[10]{}\mathrel{=}\VarId{Int}\mbox{\onelinecomment  ``node index''}{}\<[E]%
\ColumnHook
\end{hscode}\resethooks
The port index type \ensuremath{\VarId{PI}}
will be used also in actual nets, while
the node index type \ensuremath{\VarId{NI}}
is needed only for right-hand side nodes
in descriptions and during creation.
%
We arbitrarily call the two nodes engaged in an interaction ``source'' and ``target'';
the ``source'' interface consists of the auxiliary ports of the node with the ``function''
label with negative principal port, and the ``target'' interface
consists of the auxiliary ports of the ``constructor'' node
with positive principal port.
The following data type serves to identify all ports in a rule's right-hand side
(the ``\ensuremath{\mathbin{!}}'' specifies strict constructor argument positions
 for efficiency):

\begin{hscode}\SaveRestoreHook
\column{B}{@{}>{\hspre}l<{\hspost}@{}}%
\column{3}{@{}>{\hspre}c<{\hspost}@{}}%
\column{3E}{@{}l@{}}%
\column{6}{@{}>{\hspre}l<{\hspost}@{}}%
\column{20}{@{}>{\hspre}l<{\hspost}@{}}%
\column{29}{@{}>{\hspre}l<{\hspost}@{}}%
\column{E}{@{}>{\hspre}l<{\hspost}@{}}%
\>[B]{}\mathbf{data}\;\VarId{PortTargetDescription}{}\<[E]%
\\
\>[B]{}\hsindent{3}{}\<[3]%
\>[3]{}\mathrel{=}{}\<[3E]%
\>[6]{}\VarId{SourcePort}{}\<[20]%
\>[20]{}\mathbin{!}\VarId{PI}{}\<[E]%
\\
\>[B]{}\hsindent{3}{}\<[3]%
\>[3]{}\mid {}\<[3E]%
\>[6]{}\VarId{InternalPort}{}\<[20]%
\>[20]{}\mathbin{!}\VarId{NI}\mathbin{!}\VarId{PI}{}\<[29]%
\>[29]{}\mbox{\onelinecomment  node, port}{}\<[E]%
\\
\>[B]{}\hsindent{3}{}\<[3]%
\>[3]{}\mid {}\<[3E]%
\>[6]{}\VarId{TargetPort}{}\<[20]%
\>[20]{}\mathbin{!}\VarId{PI}\;{}\<[29]%
\>[29]{}\mathbf{deriving}\;(\VarId{Eq},\VarId{Ord},\VarId{Show}){}\<[E]%
\ColumnHook
\end{hscode}\resethooks
%
Therefore, each RHS node is described by its label
and by the connections of \emph{all} its ports:

\begin{hscode}\SaveRestoreHook
\column{B}{@{}>{\hspre}l<{\hspost}@{}}%
\column{3}{@{}>{\hspre}l<{\hspost}@{}}%
\column{23}{@{}>{\hspre}l<{\hspost}@{}}%
\column{42}{@{}>{\hspre}l<{\hspost}@{}}%
\column{E}{@{}>{\hspre}l<{\hspost}@{}}%
\>[B]{}\mathbf{data}\;\VarId{NodeDescription}\;\VarId{nLab}\mathrel{=}\VarId{NodeDescription}{}\<[E]%
\\
\>[B]{}\hsindent{3}{}\<[3]%
\>[3]{}\{\mskip1.5mu \VarId{nLab}{}\<[23]%
\>[23]{}\mathbin{::}{}\<[42]%
\>[42]{}\mathbin{!}\VarId{nLab}{}\<[E]%
\\
\>[B]{}\hsindent{3}{}\<[3]%
\>[3]{},\VarId{portDescriptions}{}\<[23]%
\>[23]{}\mathbin{::}\mbox{\enskip\{-\# UNPACK  \#-\}\enskip}{}\<[42]%
\>[42]{}\mathbin{!}(\VarId{Vector}\;\VarId{PortTargetDescription}){}\<[E]%
\\
\>[B]{}\hsindent{3}{}\<[3]%
\>[3]{}\mskip1.5mu\}{}\<[E]%
\ColumnHook
\end{hscode}\resethooks
A \ensuremath{\VarId{NetDescription}} is intended as description of the RHS of interaction rules:

\begin{hscode}\SaveRestoreHook
\column{B}{@{}>{\hspre}l<{\hspost}@{}}%
\column{3}{@{}>{\hspre}l<{\hspost}@{}}%
\column{13}{@{}>{\hspre}l<{\hspost}@{}}%
\column{E}{@{}>{\hspre}l<{\hspost}@{}}%
\>[B]{}\mathbf{data}\;\VarId{NetDescription}\;\VarId{nLab}\mathrel{=}\VarId{NetDescription}{}\<[E]%
\\
\>[B]{}\hsindent{3}{}\<[3]%
\>[3]{}\{\mskip1.5mu \VarId{source}{}\<[13]%
\>[13]{}\mathbin{::}\mbox{\enskip\{-\# UNPACK  \#-\}\enskip}\mathbin{!}(\VarId{Vector}\;\VarId{PortTargetDescription}){}\<[E]%
\\
\>[B]{}\hsindent{3}{}\<[3]%
\>[3]{},\VarId{target}{}\<[13]%
\>[13]{}\mathbin{::}\mbox{\enskip\{-\# UNPACK  \#-\}\enskip}\mathbin{!}(\VarId{Vector}\;\VarId{PortTargetDescription}){}\<[E]%
\\
\>[B]{}\hsindent{3}{}\<[3]%
\>[3]{},\VarId{nodes}{}\<[13]%
\>[13]{}\mathbin{::}\mbox{\enskip\{-\# UNPACK  \#-\}\enskip}\mathbin{!}(\VarId{Vector}\;(\VarId{NodeDescription}\;\VarId{nLab})){}\<[E]%
\\
\>[B]{}\hsindent{3}{}\<[3]%
\>[3]{}\mskip1.5mu\}{}\<[E]%
\ColumnHook
\end{hscode}\resethooks
%
%
A \emph{language} for interaction nets
consists of a type of node labels
together with arity and polarity information defining \emph{all} ports
for each node label, and for any ``function'' node label \ensuremath{\VarId{f}}
and any ``constructor'' node label \ensuremath{\VarId{c}} that can occur as ``argument'' to \ensuremath{\VarId{f}}
a rule, specified by a right-hand side \ensuremath{\VarId{ruleRHS}\;\VarId{f}\;\VarId{c}},
which needs to be a net description having a source compatible with the
\emph{auxiliary} ports of \ensuremath{\VarId{f}}, and a target compatible with the auxiliary ports of \ensuremath{\VarId{c}}.

\begin{hscode}\SaveRestoreHook
\column{B}{@{}>{\hspre}l<{\hspost}@{}}%
\column{32}{@{}>{\hspre}l<{\hspost}@{}}%
\column{44}{@{}>{\hspre}l<{\hspost}@{}}%
\column{E}{@{}>{\hspre}l<{\hspost}@{}}%
\>[B]{}\mathbf{data}\;\VarId{INetLang}\;\VarId{nLab}\mathrel{=}\VarId{INetLang}\;{}\<[32]%
\>[32]{}\{\mskip1.5mu \VarId{polarity}{}\<[44]%
\>[44]{}\mathbin{::}\mathbin{!}(\VarId{nLab}\to \VarId{Vector}\;\VarId{Polarity}){}\<[E]%
\\
\>[32]{},\VarId{ruleRHS}{}\<[44]%
\>[44]{}\mathbin{::}\mathbin{!}(\VarId{nLab}\to \VarId{nLab}\to \VarId{NetDescription}\;\VarId{nLab}){}\<[E]%
\\
\>[32]{}\mskip1.5mu\}{}\<[E]%
\ColumnHook
\end{hscode}\resethooks

\section{Interaction Net Reduction}\sectlabel{INet.Polar.Reduce}

\begin{ModuleHead}
\begin{hscode}\SaveRestoreHook
\column{B}{@{}>{\hspre}l<{\hspost}@{}}%
\column{27}{@{}>{\hspre}l<{\hspost}@{}}%
\column{E}{@{}>{\hspre}l<{\hspost}@{}}%
\>[B]{}\mbox{\enskip\{-\# LANGUAGE ScopedTypeVariables, RecursiveDo  \#-\}\enskip}{}\<[E]%
\\
\>[B]{}\mathbf{module}\;\VarId{\VarId{INet}.\VarId{Polar}.Reduce}\;\mathbf{where}{}\<[E]%
\\[\blanklineskip]%
\>[B]{}\mathbf{import}\;\VarId{\VarId{INet}.Polarity}\;(\VarId{Polarity}\;(\mathinner{\ldotp\ldotp})){}\<[E]%
\\
\>[B]{}\mathbf{import}\;\VarId{\VarId{INet}.\VarId{Polar}.INet}{}\<[E]%
\\
\>[B]{}\mathbf{import}\;\VarId{\VarId{INet}.Description}{}\<[E]%
\\
\>[B]{}\mathbf{import}\;\VarId{\VarId{INet}.\VarId{Utils}.MVar}{}\<[E]%
\\[\blanklineskip]%
\>[B]{}\mathbf{import}\;\VarId{\VarId{INet}.\VarId{Utils}.Vector}\;{}\<[27]%
\>[27]{}(\VarId{Vector},(\mathbin{!?}),\VarId{atErr},\VarId{bounds},(\mathbin{!})){}\<[E]%
\\
\>[B]{}\mathbf{import}\;\VarId{qualified}\;\VarId{\VarId{INet}.\VarId{Utils}.Vector}\;\VarId{as}\;\VarId{V}{}\<[E]%
\\[\blanklineskip]%
\>[B]{}\mathbf{import}\;\VarId{\VarId{Control}.Concurrent}{}\<[E]%
\\
\>[B]{}\mathbf{import}\;\VarId{\VarId{Control}.\VarId{Concurrent}.MVar}{}\<[E]%
\ColumnHook
\end{hscode}\resethooks
\end{ModuleHead}

The main purpose of the function \ensuremath{\VarId{replaceNet}} is to implement
the instantiation part of the rule application step.
It is a separate function because it also serves the
secondary purpose of constructing the start net.

The function \ensuremath{\VarId{replaceNet}} takes as arguments a \ensuremath{\VarId{NetDescription}}
(defined in \sectref{INet.Description})
for the rule's RHS,
and arrays \ensuremath{\VarId{src}} and \ensuremath{\VarId{trg}} containing the non-principal connections
of the two nodes of the image of rule's LHS in the mutable net representation
(\sectref{INet.Polar.INet}) of the run-time state.

The \ensuremath{\VarId{mdo}} is a ``recursive do'' as introduced by
\citet{Erkoek-Launchbury-2002},
and the use here essentially corresponds to the imperative
programming pattern of allocating an array of uninitialised cells,
and creating references to the array cells
possibly before initialising them.
(Functions prefix with ``\textsf{V.}'' operate on \ensuremath{\VarId{Vector}}s.)

%
\savecolumns
\begin{hscode}\SaveRestoreHook
\column{B}{@{}>{\hspre}l<{\hspost}@{}}%
\column{5}{@{}>{\hspre}l<{\hspost}@{}}%
\column{13}{@{}>{\hspre}l<{\hspost}@{}}%
\column{17}{@{}>{\hspre}l<{\hspost}@{}}%
\column{20}{@{}>{\hspre}l<{\hspost}@{}}%
\column{25}{@{}>{\hspre}l<{\hspost}@{}}%
\column{27}{@{}>{\hspre}l<{\hspost}@{}}%
\column{33}{@{}>{\hspre}l<{\hspost}@{}}%
\column{39}{@{}>{\hspre}l<{\hspost}@{}}%
\column{59}{@{}>{\hspre}l<{\hspost}@{}}%
\column{E}{@{}>{\hspre}l<{\hspost}@{}}%
\>[B]{}\VarId{replaceNet}{}\<[13]%
\>[13]{}\mathbin{::}\VarId{forall}\;\VarId{nLab}\mathbin{\circ}\VarId{INetLang}\;\VarId{nLab}\to \VarId{NetDescription}\;\VarId{nLab}{}\<[E]%
\\
\>[13]{}\to \VarId{Ports}\;\VarId{nLab}\to \VarId{Ports}\;\VarId{nLab}\to \VarId{IO}\;(){}\<[E]%
\\
\>[B]{}\VarId{replaceNet}\;\VarId{lang}\;\VarId{descr}\;\VarId{src}\;\VarId{trg}\mathrel{=}\VarId{mdo}{}\<[E]%
\\
\>[B]{}\hsindent{5}{}\<[5]%
\>[5]{}\VarId{nps}\leftarrow \mathbf{let}\;{}\<[17]%
\>[17]{}\VarId{mkNode}\;{}\<[25]%
\>[25]{}(\VarId{NodeDescription}\;\VarId{lab}\;\VarId{pds})\mathrel{=}\mathbf{do}{}\<[E]%
\\
\>[25]{}\VarId{ps}\leftarrow \VarId{\VarId{V}.zipWithM}\;\VarId{mkPort}\;(\VarId{polarity}\;\VarId{lang}\;\VarId{lab})\;\VarId{pds}{}\<[E]%
\\
\>[25]{}\VarId{return}\;{}\<[33]%
\>[33]{}(\VarId{Node}\;\{\mskip1.5mu \VarId{label}\mathrel{=}\VarId{lab},\VarId{ports}\mathrel{=}\VarId{\VarId{V}.tail}\;\VarId{ps}\mskip1.5mu\}{}\<[E]%
\\
\>[33]{},\VarId{\VarId{V}.head}\;\VarId{ps}{}\<[E]%
\\
\>[33]{}){}\<[E]%
\\
\>[17]{}\hsindent{3}{}\<[20]%
\>[20]{}\mathbf{where}\;{}\<[27]%
\>[27]{}\VarId{mkPort}\;\VarId{Pos}\;{}\<[39]%
\>[39]{}(\VarId{InternalPort}\;\anonymous \;\anonymous ){}\<[59]%
\>[59]{}\mathrel{=}\VarId{fmap}\;(\VarId{Port}\;\VarId{Pos})\;\VarId{newEmptyMVar}{}\<[E]%
\\
\>[27]{}\VarId{mkPort}\;\anonymous \;{}\<[39]%
\>[39]{}\VarId{ptd}{}\<[59]%
\>[59]{}\mathrel{=}\VarId{return}\;(\VarId{portTarget}\;\VarId{ptd}){}\<[E]%
\\
\>[17]{}\mathbf{in}\;\VarId{\VarId{V}.mapM}\;\VarId{mkNode}\;(\VarId{nodes}\;\VarId{descr}){}\<[E]%
\ColumnHook
\end{hscode}\resethooks
The first step above creates \ensuremath{\VarId{descr}} image nodes,
taking over interface ports from \ensuremath{\VarId{src}} and \ensuremath{\VarId{trg}},
creating new internal connections at positive ports,
and lazily connecting negative ports with internal connections
located via the function \ensuremath{\VarId{portTarget}} defined below.

Note that the prose explanations here
are interspersed within the scope of the \ensuremath{\VarId{mdo}} above,
since all code before the definition of \ensuremath{\VarId{reduce}} below remains indented below the \ensuremath{\VarId{mdo}}.
\restorecolumns
\begin{hscode}\SaveRestoreHook
\column{B}{@{}>{\hspre}l<{\hspost}@{}}%
\column{5}{@{}>{\hspre}l<{\hspost}@{}}%
\column{10}{@{}>{\hspre}l<{\hspost}@{}}%
\column{12}{@{}>{\hspre}l<{\hspost}@{}}%
\column{34}{@{}>{\hspre}l<{\hspost}@{}}%
\column{47}{@{}>{\hspre}l<{\hspost}@{}}%
\column{E}{@{}>{\hspre}l<{\hspost}@{}}%
\>[5]{}\mathbf{let}\;{}\<[10]%
\>[10]{}\VarId{portTarget}\mathbin{::}\VarId{PortTargetDescription}\to \VarId{Port}\;\VarId{nLab}{}\<[E]%
\\
\>[10]{}\VarId{portTarget}\;(\VarId{SourcePort}\;{}\<[34]%
\>[34]{}\VarId{i})\mathrel{=}\VarId{atErr}\;\text{\tt \char34 portTarget:~SourcePort~S\char34}\;\VarId{src}\;(\VarId{pred}\;\VarId{i}){}\<[E]%
\\
\>[10]{}\VarId{portTarget}\;(\VarId{TargetPort}\;{}\<[34]%
\>[34]{}\VarId{i})\mathrel{=}\VarId{atErr}\;\text{\tt \char34 portTarget:~TargetPort~S\char34}\;\VarId{trg}\;(\VarId{pred}\;\VarId{i}){}\<[E]%
\\
\>[10]{}\VarId{portTarget}\;(\VarId{InternalPort}\;\VarId{n}\;\VarId{i})\mathrel{=}\mathbf{let}\;{}\<[47]%
\>[47]{}\VarId{e}\mathrel{=}\text{\tt \char34 portTarget:~InternalPort~\char34}{}\<[E]%
\\
\>[47]{}(\VarId{n'},\VarId{pp})\mathrel{=}\VarId{atErr}\;\VarId{e}\;\VarId{nps}\;\VarId{n}{}\<[E]%
\\
\>[10]{}\hsindent{2}{}\<[12]%
\>[12]{}\mathbf{in}\;\VarId{opPort}\;(\mathbf{if}\;\VarId{i}\equiv \mathrm{0}\;\mathbf{then}\;\VarId{pp}\;\mathbf{else}\;\VarId{atErr}\;(\VarId{e}\plus \VarId{shows}\;\VarId{n}\;\text{\tt \char34 ~S\char34})\;(\VarId{ports}\;\VarId{n'})\;(\VarId{pred}\;\VarId{i})){}\<[E]%
\ColumnHook
\end{hscode}\resethooks
We traverse the newly created nodes and ``connect'' their principal ports.
\restorecolumns
\begin{hscode}\SaveRestoreHook
\column{B}{@{}>{\hspre}l<{\hspost}@{}}%
\column{5}{@{}>{\hspre}l<{\hspost}@{}}%
\column{8}{@{}>{\hspre}l<{\hspost}@{}}%
\column{10}{@{}>{\hspre}l<{\hspost}@{}}%
\column{14}{@{}>{\hspre}l<{\hspost}@{}}%
\column{E}{@{}>{\hspre}l<{\hspost}@{}}%
\>[5]{}\mathbf{let}\;{}\<[10]%
\>[10]{}\VarId{doNode}\;(\VarId{n}\textsf{@}(\VarId{Node}\;\VarId{lab}\;\VarId{prts}),\VarId{Port}\;\VarId{pl}\;\VarId{c})\mathrel{=}\mathbf{case}\;\VarId{pl}\;\mathbf{of}{}\<[E]%
\\
\>[10]{}\hsindent{4}{}\<[14]%
\>[14]{}\VarId{Neg}\to \VarId{forkIO}\;(\VarId{reduce}\;\VarId{lang}\;(\VarId{ruleRHS}\;\VarId{lang}\;\VarId{lab})\;\VarId{c}\;\VarId{prts})\sequ \VarId{return}\;(){}\<[E]%
\\
\>[10]{}\hsindent{4}{}\<[14]%
\>[14]{}\VarId{Pos}\to \VarId{putMVar}\;\VarId{c}\;\VarId{n}{}\<[E]%
\\
\>[5]{}\hsindent{3}{}\<[8]%
\>[8]{}\mathbf{in}\;\VarId{\VarId{V}.mapM\char95 }\;\VarId{doNode}\;\VarId{nps}{}\<[E]%
\ColumnHook
\end{hscode}\resethooks
For source and target ports, we only need to take care of short-circuits:
\restorecolumns
\begin{hscode}\SaveRestoreHook
\column{B}{@{}>{\hspre}l<{\hspost}@{}}%
\column{5}{@{}>{\hspre}l<{\hspost}@{}}%
\column{8}{@{}>{\hspre}l<{\hspost}@{}}%
\column{10}{@{}>{\hspre}l<{\hspost}@{}}%
\column{12}{@{}>{\hspre}l<{\hspost}@{}}%
\column{14}{@{}>{\hspre}l<{\hspost}@{}}%
\column{15}{@{}>{\hspre}l<{\hspost}@{}}%
\column{27}{@{}>{\hspre}l<{\hspost}@{}}%
\column{29}{@{}>{\hspre}l<{\hspost}@{}}%
\column{48}{@{}>{\hspre}l<{\hspost}@{}}%
\column{52}{@{}>{\hspre}l<{\hspost}@{}}%
\column{55}{@{}>{\hspre}l<{\hspost}@{}}%
\column{62}{@{}>{\hspre}l<{\hspost}@{}}%
\column{E}{@{}>{\hspre}l<{\hspost}@{}}%
\>[5]{}\mathbf{let}\;{}\<[10]%
\>[10]{}\VarId{doIfacePort}\;(\VarId{Port}\;\VarId{Pos}\;\VarId{c})\;\VarId{ptd}\mathrel{=}\VarId{return}\;(){}\<[52]%
\>[52]{}\mbox{\onelinecomment  will be done from the other side if necessary}{}\<[E]%
\\
\>[10]{}\VarId{doIfacePort}\;(\VarId{Port}\;\VarId{Neg}\;\VarId{c})\;\VarId{ptd}\mathrel{=}\mathbf{let}\;{}\<[52]%
\>[52]{}\mbox{\onelinecomment  original port of the LHS node}{}\<[E]%
\\
\>[10]{}\hsindent{4}{}\<[14]%
\>[14]{}\VarId{Port}\;\VarId{\char95 pl'}\;\VarId{c'}\mathrel{=}\VarId{portTarget}\;\VarId{ptd}{}\<[52]%
\>[52]{}\mbox{\onelinecomment  connecting port in image of RHS}{}\<[E]%
\\
\>[10]{}\hsindent{2}{}\<[12]%
\>[12]{}\mathbf{in}\;\mathbf{if}\;\VarId{c}\equiv \VarId{c'}\;{}\<[27]%
\>[27]{}\mathbf{then}\;\VarId{return}\;(){}\<[52]%
\>[52]{}\mbox{\onelinecomment  empty cycle}{}\<[E]%
\\
\>[27]{}\mathbf{else}\;\mathbf{case}\;\VarId{ptd}\;\mathbf{of}{}\<[E]%
\\
\>[27]{}\hsindent{2}{}\<[29]%
\>[29]{}\VarId{InternalPort}\;\VarId{n}\;\VarId{i'}{}\<[48]%
\>[48]{}\to \VarId{return}\;(){}\<[62]%
\>[62]{}\mbox{\onelinecomment  already dealt with}{}\<[E]%
\\
\>[27]{}\hsindent{2}{}\<[29]%
\>[29]{}\anonymous {}\<[48]%
\>[48]{}\to \mathbf{do}\;{}\<[55]%
\>[55]{}\VarId{forkIO}\;(\VarId{moveMVar}\;\VarId{c}\;\VarId{c'}){}\<[E]%
\\
\>[55]{}\VarId{return}\;(){}\<[E]%
\\
\>[5]{}\hsindent{3}{}\<[8]%
\>[8]{}\mathbf{in}\;\mathbf{do}\;{}\<[15]%
\>[15]{}\VarId{\VarId{V}.zipWithM\char95 }\;\VarId{doIfacePort}\;\VarId{src}\mathbin{\$}\VarId{source}\;\VarId{descr}{}\<[E]%
\\
\>[15]{}\VarId{\VarId{V}.zipWithM\char95 }\;\VarId{doIfacePort}\;\VarId{trg}\mathbin{\$}\VarId{target}\;\VarId{descr}{}\<[E]%
\ColumnHook
\end{hscode}\resethooks
%
Whenever a function node is created,
i.e., a node with positive principal port,
a \ensuremath{\VarId{reduce}} thread is started (via \ensuremath{\VarId{forkIO}}).
This thread waits on the connection (\ensuremath{\VarId{pconn}}) between the principal ports of the rule
until this contains the constructor node
(the principal port of which has positive polarity).
The array \ensuremath{\VarId{src}} contains the auxiliary ports of the function node
(the principal port of which has negative polarity).
\begin{hscode}\SaveRestoreHook
\column{B}{@{}>{\hspre}l<{\hspost}@{}}%
\column{4}{@{}>{\hspre}l<{\hspost}@{}}%
\column{E}{@{}>{\hspre}l<{\hspost}@{}}%
\>[B]{}\VarId{reduce}\mathbin{::}\VarId{INetLang}\;\VarId{nLab}\to (\VarId{nLab}\to \VarId{NetDescription}\;\VarId{nLab})\to \VarId{Conn}\;\VarId{nLab}\to \VarId{Ports}\;\VarId{nLab}\to \VarId{IO}\;(){}\<[E]%
\\
\>[B]{}\VarId{reduce}\;\VarId{lang}\;\VarId{rules}\;\VarId{pconn}\;\VarId{src}\mathrel{=}\mathbf{do}{}\<[E]%
\\
\>[B]{}\hsindent{4}{}\<[4]%
\>[4]{}\VarId{Node}\;\VarId{clab}\;\VarId{trg}\leftarrow \VarId{takeMVar}\;\VarId{pconn}{}\<[E]%
\\
\>[B]{}\hsindent{4}{}\<[4]%
\>[4]{}\VarId{replaceNet}\;\VarId{lang}\;(\VarId{rules}\;\VarId{clab})\;\VarId{src}\;\VarId{trg}{}\<[E]%
\ColumnHook
\end{hscode}\resethooks


}

\section{Reading \texttt{.inet} Files}\sectlabel{RunInets}

The \Inets{} project led by Ian Mackie
has implemented the only publicly available general implementation
of interaction nets, the compiler \cite{INets}
for the interaction net programming language ``\Inets{}''.
This language was introduced by Mackie \cite{Mackie-2005},
with the core of the \Inets{} implementation described in \cite{Hassan-Mackie-Sato-2009}.

We implemented a front-end to our interaction net reduction system
for the core sublanguage of \Inets{},
leaving out in particular
the extension of nested
pattern matching described in \cite{Hassan-Mackie-Sato-2010},
and generic rules and variadic agents.

Since our system depends on polarity for its directed implementation
of connections, but \Inets{} has no concept of polarity,
we adopted the convention that the first-mentioned agent
of each rule has negative principal port (that is, is considered as a
function),
and the second agent has positive principal port (constructor).
This convention is adopted in most of the \Inets{} examples anyways;
only two rules in \textsf{fibonacci.inet} had been written the other way
around.
From this starting point we attempt to deduce the polarities of all other ports;
for the examples accessible to us so far, we only needed to add
a single additional heuristic:
A function for which all other ports except one are known to have
negative polarity is assumed to have positive polarity on the last
port.
(Unfortunately the $\lambda$-calculus evaluator \textsf{yale.inet}
\cite{Mackie-1998} is defined in a way that does not allow a
consistent assignment of polarities.)

\Inets{} supports ``parameters'', that is, agent attributes
of the primitive types \ensuremath{\VarId{int}}, \ensuremath{\VarId{bool}}, \ensuremath{\VarId{float}}, \ensuremath{\VarId{char}}, and \ensuremath{\VarId{String}}.
The description in \cite{Hassan-Mackie-Sato-2009} suggests that
only a single parameter is allowed per agent;
our implementation allows arbitrary numbers, but expects the number
and types of attributes to be determined by the agent label.
We also interpret type \ensuremath{\VarId{int}} as Haskell's arbitrary-precision
\ensuremath{\VarId{Integer}} type.
Our current interpreting implementation uses a parameterised agent
label type:
\begin{hscode}\SaveRestoreHook
\column{B}{@{}>{\hspre}l<{\hspost}@{}}%
\column{E}{@{}>{\hspre}l<{\hspost}@{}}%
\>[B]{}\mathbf{data}\;\VarId{NLab}\;\VarId{arg}\mathrel{=}\VarId{NLab}\;\{\mskip1.5mu \VarId{nLabName}\mathbin{::}\VarId{Name},\VarId{nLabAttrs}\mathbin{::}[\mskip1.5mu \VarId{arg}\mskip1.5mu]\mskip1.5mu\}{}\<[E]%
\ColumnHook
\end{hscode}\resethooks
When reading a \textsf{.inet} file,
the nets on the rule RHSs are translated into
\ensuremath{\VarId{NetDescription}\;(\VarId{NLab}\;\VarId{Expression})} and stored in a finite map for
lookup by the rule LHS agent label pair;
in the run-time net, agent labels of type \ensuremath{\VarId{NLab}\;\VarId{Value}} are used,
and the variable bindings induced by the attributes of the interacting
nodes are used at the time of rule application
to evaluate the expressions in the RHSs
(and the condition expressions for the conditional structure of \Inets{} RHSs).

\Inets{} modules can contain global variables, which are used
in the examples to implement reduction counts etc.;
since in a parallel implementation such global variables would require
synchronisation (and thus would destroy the independence of parallel
reduction), we did not implement any feature related to global variables.

\section{Benchmarks}\sectlabel{Benchmarks}

\noindent
For our first examples, we use a cascading recursion for calculating Fibonacci
numbers, and the Ackermann function, both computing with unary natural
numbers constructed from zero \ensuremath{\VarId{Z}} and the unary successor
constructor \ensuremath{\VarId{S}}:

\noindent
\strut
\begin{minipage}{0.4\columnwidth}\small
\begin{hscode}\SaveRestoreHook
\column{B}{@{}>{\hspre}l<{\hspost}@{}}%
\column{E}{@{}>{\hspre}l<{\hspost}@{}}%
\>[B]{}\VarId{fib}\;\mathrm{0}\mathrel{=}\mathrm{0}{}\<[E]%
\\
\>[B]{}\VarId{fib}\;(\VarId{S}\;\VarId{n})\mathrel{=}\VarId{fibAux}\;\VarId{n}{}\<[E]%
\\
\>[B]{}\VarId{fibAux}\;\mathrm{0}\mathrel{=}\mathrm{1}{}\<[E]%
\\
\>[B]{}\VarId{fibAux}\;(\VarId{S}\;\VarId{n})\mathrel{=}\VarId{fib}\;\VarId{n}\mathbin{+}\VarId{fibAux}\;\VarId{n}{}\<[E]%
\ColumnHook
\end{hscode}\resethooks
\end{minipage}
\hfill
\begin{minipage}{0.5\columnwidth}\small
\begin{hscode}\SaveRestoreHook
\column{B}{@{}>{\hspre}l<{\hspost}@{}}%
\column{E}{@{}>{\hspre}l<{\hspost}@{}}%
\>[B]{}\VarId{ack}\;\mathrm{0}\;\VarId{n}\mathrel{=}\VarId{S}\;\VarId{n}{}\<[E]%
\\
\>[B]{}\VarId{ack}\;(\VarId{S}\;\VarId{m})\;\VarId{n}\mathrel{=}\VarId{ackAux}\;\VarId{m}\;\VarId{n}{}\<[E]%
\\
\>[B]{}\VarId{ackAux}\;\VarId{m}\;\mathrm{0}\mathrel{=}\VarId{ack}\;\VarId{m}\;\mathrm{1}{}\<[E]%
\\
\>[B]{}\VarId{ackAux}\;\VarId{m}\;(\VarId{S}\;\VarId{n})\mathrel{=}\VarId{ack}\;\VarId{m}\;(\VarId{ack}\;(\VarId{S}\;\VarId{m})\;\VarId{n}){}\<[E]%
\ColumnHook
\end{hscode}\resethooks
\end{minipage}

\noindent
These rules were directly encoded using \ensuremath{\VarId{NetDescription}}s (see
\sectref{INet.Description});
we will refer to these implementations now as \ensuremath{\VarId{fibND}} and \ensuremath{\VarId{ackND}}.

We timed the actual code of \sectref{Implementation}
on a six-core 2.8GHz Phenom 2 with 16GB main memory;
our implementation achieved the timings in Table \ref{NDTable},
where the GHC run-time system is
instructed by ``\ensuremath{\mathbin{-}\VarId{N}}$k$'' to use $k$ cores for parallel processing.
The user-space time of a Haskell process is divided into ``mutation''
time
and garbage collection time. The run-time system can be made to report
these times and further information;
in Tables \ref{NDTable} and \ref{inetTable} we include, after the elapsed time for each
process (which is the ``real'' time as reported by ``\texttt{time}''
BASH built-in),
the ``allocation rate'', which measures how many megabytes are allocated
on the Haskell heap per second of mutation time,
and the ``productivity'', which is the result of dividing the mutation
time by the elapsed time.
For example, a productivity of 240\% for a three-core (``-N3'') run
means that each core spent on average 20\% of its time on garbage
collection, since $240\% + 3 \times 20\% = 300\%$.
The last column in each of the groups for ``-N2'' to ``-N6''
contains the speedup over single-core execution.

By default, the GHC run-time system starts execution with a small heap
and grows it by relatively small increments on demand;
we indicate use of this this default setting by ``\defaultHeap'' in the third column
(``heap''). Where a size is specified in this column,
this size was given to the run-time system as fixed heaps size
(with options -H \emph{and} -M).

\begin{table}[ht]
\centerline{\scalebox{1.04}{\tiny
\setlength{\tabcolsep}{0.2em}
\begin{tabular}{|l||r||r||r|r|r||r|r|r|>{\bfseries}r||r|r|r|>{\bfseries}r||r|r|r|>{\bfseries}r||r|r|r|>{\bfseries}r||r|r|r|>{\bfseries}r|}
\hline
& & & \multicolumn{23}{c|}{time (s)
  \qquad$\mid$\qquad
  allocation~rate (MB per mutation second)
  \qquad$\mid$\qquad
  productivity (\% of elapsed)
  \qquad$\mid$\qquad
  \textbf{speedup}}
\\
expr. & result & heap
   & \multicolumn{3}{@{\kern-0.1ex}|c|@{\kern0.8ex}|}{-N{\bfseries 1}}
   & \multicolumn{4}{@{\kern-0.1ex}|c|@{\kern0.8ex}|}{-N{\bfseries 2}}
   & \multicolumn{4}{@{\kern-0.1ex}|c|@{\kern0.8ex}|}{-N{\bfseries 3}}
   & \multicolumn{4}{@{\kern-0.1ex}|c|@{\kern0.8ex}|}{-N{\bfseries 4}}
   & \multicolumn{4}{@{\kern-0.1ex}|c|@{\kern0.8ex}|}{-N{\bfseries 5}}
   & \multicolumn{4}{@{\kern-0.1ex}|c|}{-N{\bfseries 6}}
\\
\hline
ackND 3 6 & 509 & \defaultHeap
  & 1.078 & 1699 & 48
  & 0.630 & 1189 & 118 & 1.71
  & 0.483 & 952 & 192 & 2.23
  & 0.431 & 779 & 264 & 2.50
  & 0.427 & 611 & 340 & \best{2.52}
  & 0.426 & 506 & 411 & \best{2.53}
  \\
ackND 3 6 & 509 & 2M
  & 1.004 & 1716 & 51
  & 0.633 & 1188 & 118 & 1.59
  & 0.489 & 960 & 189 & 2.05
  & 0.452 & 738 & 266 & 2.22
  & 0.411 & 647 & 334 & \best{2.44}
  & 0.421 & 516 & 409 & 2.38
  \\
ackND 3 6 & 509 & 3M
  & 0.800 & 1693 & 65
  & 0.568 & 1196 & 130 & 1.41
  & 0.478 & 960 & 193 & 1.67
  & 0.445 & 741 & 269 & 1.80
  & 0.409 & 652 & 333 & \best{1.96}
  & 0.429 & 509 & 405 & 1.86
  \\
ackND 3 6 & 509 & 4M
  & 0.694 & 1678 & 76 
  & 0.504 & 1202 & 146 & 1.38
  & 0.444 & 942 & 212 & 1.56
  & 0.433 & 747 & 274 & 1.60
  & 0.405 & 646 & 339 & \best{1.71}
  & 0.425 & 512 & 407 & 1.63
  \\
ackND 3 6 & 509 & 5M
  & 0.652 & 1652 & 82
  & 0.475 & 1194 & 156 & 1.37
  & 0.415 & 948 & 225 & 1.57
  & 0.404 & 751 & 293 & 1.61
  & 0.385 & 652 & 354 & \best{1.69}
  & 0.422 & 510 & 413 & 1.55
  \\
ackND 3 6 & 509 & 6M
  & 0.647 & 1604 & 85.3
  & 0.462 & 1181 & 163 & 1.40
  & 0.400 & 945 & 235 & 1.62
  & 0.387 & 745 & 308 & 1.67
  & 0.375 & 647 & 366 & \best{1.73}
  & 0.395 & 522 & 431 & 1.64
  \\
ackND 3 6 & 509 & 7M
  & 0.644 & 1575 & 87
  & 0.459 & 1159 & 167 & 1.40
  & 0.389 & 943 & 242 & 1.66
  & 0.382 & 739 & 315 & 1.69
  & \best{0.361} & 652 & 377 & \best{1.78}
  & 0.395 & 521 & 431 & 1.63
  \\
ackND 3 6 & 509 & 8M
  & 0.659 & 1521 & 88
  & 0.472 & 1108 & 170 & 1.40
  & 0.396 & 918 & 244 & 1.66
  & 0.384 & 727 & 319 & 1.72
  & 0.363 & 640 & 383 & \best{1.81}
  & 0.388 & 511 & 447 & 1.69
  \\
ackND 3 6 & 509 & 9M
  & 0.675 & 1469 & 89
  & 0.482 & 1070 & 172 & 1.40
  & 0.411 & 873 & 248 & 1.64
  & 0.393 & 705 & 321 & 1.72
  & 0.374 & 615 & 386 & \best{1.80}
  & 0.392 & 501 & 452 & 1.72
  \\
ackND 3 6 & 509 & 10M
  & 0.686 & 1437 & 90
  & 0.485 & 1061 & 172 & 1.41
  & 0.420 & 846 & 250 & 1.63
  & 0.404 & 678 & 324 & 1.70
  & 0.379 & 600 & 391 & \best{1.81}
  & 0.399 & 489 & 456 & 1.72
\\
ackND 3 6 & 509 & 0.1G
  & 0.749 & 1305 & 92
  & 0.522 & 982 & 175 & 1.43
  & 0.445 & 796 & 253 & 1.68
  & 0.430 & 635 & 329 & 1.74
  & 0.416 & 544 & 397 & \best{1.80}
  & 0.435 & 452 & 458 & 1.72
  \\
\hline
ackND 3 7 & 1021 & \defaultHeap
  & 5.866 & 1676 & 36
  & 3.177 & 1185 & 94 & 1.85
  & 2.287 & 982 & 158 & 2.56
  & 1.990 & 802 & 223 & 2.95
  & 1.845 & 642 & 300 & 3.18
  & 1.771 & 547 & 367 & \best{3.31}
\\
ackND 3 7 & 1021 & 6M
  & 3.335 & 1585 & 67
  & 2.288 & 1181 & 131 & 1.46
  & 1.877 & 979 & 193 & 1.78
  & 1.815 & 764 & 256 & 1.84
  & 1.661 & 681 & 314 & \best{2.01}
  & 1.728 & 542 & 380 & 1.93
\\
ackND 3 7 & 1021 & 8M
  & 3.024 & 1514 & 78
  & 2.115 & 1133 & 148 & 1.43
  & 1.723 & 953 & 217 & 1.76
  & 1.666 & 759 & 281 & 1.82
  & 1.521 & 683 & 342 & \best{1.99}
  & 1.591 & 551 & 406 & 1.90
  \\
ackND 3 7 & 1021 & 9M
  & 3.000 & 1474 & 80
  & 2.076 & 1122 & 152 & 1.45
  & 1.715 & 930 & 223 & 1.75
  & 1.651 & 743 & 290 & 1.82
  & \best{1.496} & 670 & 355 & \best{2.01}
  & 1.545 & 547 & 422 & 1.94
  \\
ackND 3 7 & 1021 & 10M
  & 3.010 & 1438 & 82
  & 2.107 & 1078 & 156 & 1.43
  & 1.735 & 902 & 227 & 1.74
  & 1.638 & 729 & 298 & 1.84
  & 1.520 & 649 & 361 & \best{1.98}
  & 1.579 & 531 & 424 & 1.90
  \\
ackND 3 7 & 1021 & 20M
  & 2.932 & 1377 & 88
  & 2.022 & 1035 & 170 & 1.45
  & 1.712 & 848 & 245 & 1.71
  & 1.637 & 682 & 319 & 1.79
  & 1.539 & 592 & 390 & \best{1.91}
  & 1.579 & 490 & 460 & 1.86
\\
ackND 3 7 & 1021 & 1G
  & 3.508 & 1187 & 91
  & 2.472 & 881 & 171 & 1.41
  & 2.085 & 727 & 246 & 1.68
  & 1.971 & 590 & 319 & 1.78
  & 1.835 & 527 & 384 & \best{1.91}
  & 1.840 & 448 & 449 & 1.90
\\
\hline
ackND 3 8 & 2045 & \defaultHeap
  & 30.034 & 1557 & 30
  & 16.857 & 1138 & 74 & 1.78
  & 11.412 &  956 & 131 & 2.63
  & 9.640 & 791 & 187 & 3.12
  & 8.597 & 646 & 256 & 3.49
  & 8.061 & 550 & 322 & \best{3.73}
  \\
ackND 3 8 & 2045 & 10M
  & 17.000 & 1423 & 59
  & 11.116 & 1065 & 120 & 1.53
  & 8.802 & 899 & 180 & 1.93
  & 8.195 & 727 & 239 & 2.07
  & 7.320 & 657 & 296 & 2.32
  & 7.306 & 540 & 361 & \best{2.33}
  \\
ackND 3 8 & 2045 & 40M
  & 13.089 & 1248 & 87
  & 9.171 & 929 & 167 & 1.43
  & 7.450 & 789 & 243 & 1.76
  & 7.079 & 633 & 318 & 1.85
  & 6.503 & 564 & 389 & \best{2.01}
  & 6.539 & 473 & 461 & 2.00
  \\
ackND 3 8 & 2045 & 60M
  & 13.057 & 1228 & 89
  & 9.019 & 923 & 171 & 1.45
  & 7.373 & 778 & 248 & 1.77
  & 6.929 & 634 & 324 & 1.88
  & 6.372 & 565 & 396 & \best{2.05}
  & 6.375 & 475 & 471 & 2.05
  \\
ackND 3 8 & 2045 & 80M
  & 13.110 & 1212 & 90
  & 8.989 & 920 & 173 & 1.46
  & 7.292 & 781 & 250 & 1.80
  & 6.904 & 630 & 328 & 1.90
  & \best{6.353} & 562 & 399 & \best{2.06}
  & 6.364 & 478 & 469 & 2.06
  \\
ackND 3 8 & 2045 & 100M
  & 13.043 & 1215 & 90
  & 9.042 & 913 & 173 & 1.44
  & 7.345 & 772 & 251 & 1.76
  & 6.917 & 628 & 328 & 1.88
  & 6.372 & 559 & 400 & \best{2.05}
  & 6.376 & 475 & 470 & 2.05
  \\
ackND 3 8 & 2045 & 1G
  & 13.849 & 1154 & 90
  & 9.588 & 869 & 173 & 1.44
  & 7.824 & 737 & 250 & 1.77
  & 7.288 & 606 & 326 & 1.90
  & 6.665 & 547 & 395 & 2.08
  & 6.627 & 472 & 460 & \best{2.09}
  \\
ackND 3 8 & 2045 & 8G
  & 15.521 & 1043 & 91
  & 11.200 & 819 & 169 & 1.39
  & 9.204 & 703 & 239 & 1.69
  & 8.686 & 573 & 309 & 1.79
  & 7.947 & 523 & 370 & 1.95
  & 7.822 & 453 & 432 & \best{1.98}
\\
\hline
ackND 3 9 & 4093 & \defaultHeap
  & 141.662 & 1415 & 28
  & 85.999 & 1041 & 64 & 1.65
  & 60.032 & 904 & 105 & 2.36
  & 49.941 & 755 & 151 & 2.84
  & 42.996 & 625 & 212 & 3.29
  & 38.546 & 547 & 271 & \best{3.68}
  \\
ackND 3 9 & 4093 & 8G
  & 62.920 & 1016 & 91
  & 41.032 & 815 & 174 & 1.53
  & 32.717 & 708 & 251 & 1.92
  & 30.653 & 577 & 328 & 2.05
  & 27.727 & 526 & 398 & 2.27
  & \best{27.087} & 459 & 466 & \best{2.32}
  \\
\hline
ackND 3 10 & 8189 & 8G
  & 300.687 & 837 & 91
  & 184.245 & 716 & 174 & 1.63
  & 141.858 & 643 & 252 & 2.12
  & 128.381 & 546 & 328 & 2.34
  & 115.164 & 501 & 398 & 2.61
  & 110.754 & 447 & 464 & \best{2.71}
  \\
\hline
\hline
fibND 20 &   6765 & 1G
  & 0.513  & 667 & 94
  & 0.339  & 558 & 168  & 1.51
  & 0.283  & 444 & 232 & 1.81
  & 0.251  & 440 & 294 & 2.04
  & 0.232  & 404 & 338 & 2.21
  & 0.225  & 358 & 403 & \best{2.28}
  \\
fibND 25 &  75025 & 4G
  & 6.651 & 621 & 85
  & 4.354 & 527 & 153 & 1.53
  & 3.437 & 410 & 212 & 1.93
  & 3.097 & 410 & 276 & 2.15
  & 2.883 & 370 & 327 & 2.31
  & 2.634 & 332 & 387 & \best{2.53}
  \\
fibND 28 & 317811 & 8G
  & 32.24 & 732 & 64
  & 21.92 & 619 & 112 & 1.47
  & 16.34 & 544 & 172 & 1.97
  & 14.85 & 454 & 226 & 2.17
  & 13.41 & 412 & 278 & 2.40
  & 12.85 & 375 & 317 & \best{2.51}
  \\
fibND 30 & 832040 & 8G
  & 139.617 & 762 & 38
  & 101.619 & 647 & 62 & 1.37
  & 62.626 & 557 & 116 & 2.23
  & 56.072 & 438 & 158 & 2.49
  & 49.494 & 423 & 193 & 2.82
  & 44.478 & 371 & 247 & \best{3.14}
  \\
\hline
\end{tabular}
}}
\caption{Benchmarks for directly-programmed \ensuremath{\VarId{NetDescription}}s}\label{NDTable}
\end{table}

In general, as long as the heap is small in comparison with the space requirements
of the current run, the run-time system spends a much higher part of
its time performing garbage collection --- this manifests itself in
low ``productivity'' entries in the tables below in the rows with
small fixed heap sizes and with ``\defaultHeap''.
(The amount of space that is allocated on the heap by any given task
varies only minimally with different heap and parallelism settings.)
Not limiting the heap size (with (``-M\emph{size}'') on longer-running
tasks may lead the run-time system to use a heap that is larger than
the available physical memory, leading to drastic performance loss
dues to swapping of memory pages to peripheral storage.
For tasks that actually do use large heap space,
not fixing the start heap size (with (``-H\emph{size}'')
lets the run-time system adopt the default behaviour at the start of
the program, leading to slow-down of actually acquiring the needed
large heap.
Therefore, optimal time is typically obtained using a fixed heap size,
that is, with both -H and -M set to the same size,
which is what we adopted for our benchmarking.
(The GHC run-time system also provides finer control
over the initial heap size, and over the size of the increments;
we did not experiment with these here.)

Over its whole run-time, \ensuremath{\VarId{ackND}\;\mathrm{3}\;\mathrm{6}} allocates 880MB on the heap, and 
\ensuremath{\VarId{ackND}\;\mathrm{3}\;\mathrm{7}} allocates 3.5GB.
If such small tasks are given large heaps,
this leads to significant slow-down.
As can be seen for \ensuremath{\VarId{ackND}\;\mathrm{3}\;\mathrm{8}}, which allocates 14GB,
giving larger processes a generous fixed heap
produces a performance that is closer to the optimum
than using the default settings.

On an 8-core 16-hyperthread 2.4GHz Xeon 8870,
each of the examples we tried so far has a maximum number of cores
beyond which adding cores slows down reduction, see Table \ref{XeonTable}.
\begin{table}[h!]
\centerline{\scalebox{1.05}{\tiny
\begin{tabular}{|l||r|r|r|r|r|r|r|r||r|r|r|r|r|r|r|}
\hline
& \multicolumn{8}{c|}{time (s)} & \multicolumn{7}{|c|}{speedup factor over -N1} \\
expr. & -N1 & -N2 
   & -N5 
   & -N8 & -N9 & -N10 & -N11 & -N12
   & -N2 & -N5 & -N8 & -N9 & -N10 & -N11 & -N12\\
\hline
fib 28          & 63.581 & 40.173 
                & 22.495 
                & 19.389
                & 16.572 & 17.640 & 16.618 & 17.234
                        & 1.58 
                        & 2.83
                        & 3.28
                        & \textbf{3.84} & 3.60 &  \textbf{3.83} & 3.69
                        \\
fib 30          & 223.291 &  
                & 
               & 68.377
               & 63.488  & 58.204 & 60.160 & 62.559
                        & &  & 3.27 & 3.52 & \textbf{3.84} & 3.71 & 3.57 \\
ack 3 7        & 5.900 & 4.177 
               & 3.234 
               & 3.889
               & 3.786 & 4.042 & 4.033 & 4.170
                        & 1.41 & \textbf{1.82} & 1.52
                        & 1.56 & 1.46 & 1.46 & 1.41 \\
\hline
\end{tabular}
}}
\caption{16-core Benchmarks for directly-programmed \ensuremath{\VarId{NetDescriptions}}}\label{XeonTable}
\end{table}
%
This is an example of the effect of diminishing gains of adding
processors to a parallel workload that does not split into
a sufficient number of sufficiently large independent pieces:
The overhead of synchronisation in such a context
makes it unfeasible to profit from the computing power of added cores
beyond a task-dependent threshold.

\smallbreak
Table \ref{inetTable} contains timings for running our \ensuremath{\VarId{RunInets}}
interpreter on a collection of \Inets{} programs mostly derived
from programs in \cite{INets}
by replacing the \ensuremath{\VarId{main}} nets with larger examples.
The last two columns contain timings for running
the compiled programs using the \Inets{} compiler of \cite{INets},
and the quotient of our ``-N1'' time with this run-time.

\textsf{Ackerman.inet} from \cite{INets} uses a
(totalised) predecessor function;
\textsf{Ack.inet} is a direct translation of the rules in \ensuremath{\VarId{ackND}}.
The counts reported by the \Inets{} implementation
indicate that \textsf{Ackerman.inet} requires almost exactly 1.5 times
the number of rule applications of \textsf{Ack.inet};
\Inets{}-compiled executables and our \ensuremath{\VarId{RunInets}} take roughly 1.6 times the time.

\textsf{fib.inet} is a direct translation of our \ensuremath{\VarId{fibND}}
implementation into \Inets{}, and works, like both Ackerman functions,
on unary natural numbers constructed from \ensuremath{\VarId{S}} and \ensuremath{\VarId{Z}}.
We found that \textsf{fib.inet} performs roughly 20\% more allocation
than \ensuremath{\VarId{fibND}}, which will be due to the overhead of
transforming an \ensuremath{\VarId{Expression}}-based \ensuremath{\VarId{NetDescription}} into \ensuremath{\VarId{Value}}-based
for each rule application
(even though there are no expressions to evaluate in this example
that does not use attributes).
However, it appears that the difference in run times is, as for
\textsf{Ack.inet} versus \ensuremath{\VarId{ackND}}, much less ---
this should be due to the fact that the overhead is not slowed down
by concurrency synchronisation.

\textsf{fibonacci.inet} from \cite{INets} carries arguments and
results in node attributes, and uses implementation-provided addition
of integer attributes instead of recursing over predecessors like \ensuremath{\VarId{fibND}}.
It therefore has significantly less work to do than \ensuremath{\VarId{fibND}}.

\textsf{sort.inet} from \Inets{} is an implementation of bubble sort
on lists;
it uses an \ensuremath{\VarId{int}}-valued agent attribute to carry the list elements,
so element comparisons are performed as part of choosing the RHS
of conditional rules.
The counter results of the \Inets{} runs show that this performs
exactly $(n/2 + 1) \cdot (n+1)$ interactions for a randomly generated start list with even
length $n$.
This pattern fits some of the \ensuremath{\VarId{RunInets}} times in Table
\ref{inetTable} exactly,
while other \ensuremath{\VarId{RunInets}} times appear to
exhibit a worse asymptotic behaviour;
I suggest that this is due to the fact
that I used the same heap sizes for different sort argument sizes
instead of trying to identify respective optimal heap sizes.

\smallbreak
On the whole, on a single core, \ensuremath{\VarId{RunInets}} typically takes about 10 to 20 times
the time of the \Inets{}-compiled executables,
which is to be expected for an interpreted implementation.

\begin{table}[h!]
\centerline{\scalebox{0.94}{\tiny
\setlength{\tabcolsep}{0.13em}
\def\best#1{\fbox{\bf #1}}
\begin{tabular}{|l||r||r||r|r|r||r|r|r|>{\bfseries}r||r|r|r|>{\bfseries}r||r|r|r|>{\bfseries}r||r|r|r|>{\bfseries}r||r|r|r|>{\bfseries}r||r|r|}
\hline
& & & \multicolumn{23}{c|@{\kern0.8ex}|}{time (s), allocation~rate
  (MB/MUT-s), productivity (\% of elapsed), \textbf{speedup}}
& \multicolumn{2}{@{\kern-0.1ex}|c|@{\kern0.8ex}|}{\Inets{}}
\\
expr. & result & heap
   & \multicolumn{3}{@{\kern-0.1ex}|c|@{\kern0.8ex}|}{-N{\bfseries 1}}
   & \multicolumn{4}{@{\kern-0.1ex}|c|@{\kern0.8ex}|}{-N{\bfseries 2}}
   & \multicolumn{4}{@{\kern-0.1ex}|c|@{\kern0.8ex}|}{-N{\bfseries 3}}
   & \multicolumn{4}{@{\kern-0.1ex}|c|@{\kern0.8ex}|}{-N{\bfseries 4}}
   & \multicolumn{4}{@{\kern-0.1ex}|c|@{\kern0.8ex}|}{-N{\bfseries 5}}
   & \multicolumn{4}{@{\kern-0.1ex}|c|@{\kern0.8ex}|}{-N{\bfseries 6}}
 & time  & \rotatebox{90}{\parbox{4em}{speedup over -N1}}
\\
\hline
\hline
Ackerman 3 6 & 509 & \defaultHeap
  & 2.150  & 2100 & 38
  & 1.191 & 1440 & 100 & 1.80
  & 0.850 & 1214 & 166 & 2.53
  & 0.770 & 932 & 238 & 2.79
  & 0.722 & 758 & 312 & 2.98
  & 0.716 & 624 & 382  & \best{3.00}
& 0.125 & 17.2\\
Ackerman 3 6 & 509 & 20M
  & 1.204 & 1597 & 89
  & 0.861 & 1163 & 171 & 1.40
  & 0.744 & 920 & 251 & 1.62
  & 0.711 & 745 & 323 & 1.69
  & \best{0.700} & 618 & 396 & \best{1.72}
  & 0.740 & 497 & 466 & 1.63
& & 9.63 \\
Ackerman 3 6 & 509 & 8G
  & 2.163 & 1083 & 88
  & 1.536 & 850 & 156 & 1.41
  & 1.323 & 712 & 216 & 1.63
  & 1.236 & 608 & 268 & 1.75
  & 1.189 & 528 & 318 & \best{1.82}
  & 1.209 & 437 & 374 & 1.79
& & 17.3 \\
\hline
Ackerman 3 7 & 1021 & \defaultHeap
  & 11.432 & 2022 & 30
  & 6.251 & 1415 & 77 & 1.83
  & 4.158 & 1226 & 134 & 1.75
  & 3.668 & 936 & 200 & 3.12
  & 3.249 & 779 & 271 & 3.52
  & 3.080 & 655 & 339 & \best{3.71}
& 0.511 & 22.37 \\
Ackerman 3 7 & 1021 & 40M
  & 5.211 & 1498 & 88
  & 3.601 & 1127 & 169 & 1.48
  & 3.036 & 915 & 247 & 1.72
  & 2.799 & 759 & 322 & 1.86
  & \best{2.738} & 632 & 396 & \best{1.90}
  & 2.843 & 516 & 467 & 1.83
& & 10.2 \\
Ackerman 3 7 & 1021 & 8G
  & 8.359 & 1034 & 92
  & 5.678 & 827 & 169 & 1.47
  & 4.741 & 701 & 238 & 1.76
  & 4.291 & 609 & 302 & 1.95
  & 4.051 & 535 & 362 & 2.06
  & 4.023 & 457 & 423 & \best{2.08}
& & 16.4\\
\hline
Ackerman 3 8 & 2045 & \defaultHeap
  & 55.740 & 1899 & 26
  & 32.567 & 1329 & 63 & 1.71
  & 21.601 & 1173 & 108 & 2.58
  & 18.430 & 915 & 163 & 3.02
  & 15.819 & 772 & 225 & 3.52
  & 14.883 & 652 & 283 & \best{3.75}
& 2.050 & 27.2\\
Ackerman 3 8 & 2045 & 60M
  & 22.716 & 1408 & 86
  & 15.671 & 1062 & 165 & 1.45
  & 12.997 & 878 & 241 & 1.75
  & 11.922 & 732 & 314 & 1.91
  & 11.419 & 620 & 389 & 1.99
  & \best{11.258} & 536 & 455 & \best{2.02}
  & & 11.1 \\
Ackerman 3 8 & 2045 & 8G
  & 25.963 & 1230 & 90
  & 18.016 & 938 & 170 & 1.44
  & 14.724 & 803 & 242 & 1.76
  & 13.623 & 673 & 312 & 1.90
  & 12.904 & 585 & 379 & \best{2.01}
  & 13.042 & 490 & 446 & 1.99
& & 12.7 \\
\hline
Ackerman 3 9 & 4093 & \defaultHeap
  & 274.154 & 1583 & 25
  & 161.029 & 1207 & 57 & 1.70
  & 114.140 & 1072 & 90 & 2.42
  & 98.409 & 842 & 133 & 2.79
  & 82.098 & 725 & 185 & 3.34
  & 72.683 & 638 & 237 & \best{3.77}
& 7.167 & 38.3 \\
Ackerman 3 9 & 4093 & 100M
  & 107.704 & 1215 & 84
  & 68.144 & 1018 & 158 & 1.58
  & 54.532 & 872 & 231 & 1.98
  & 49.224 & 741 & 301 & 2.19
  & 47.475 & 618 & 375 & 2.27
  & \best{47.038} & 525 & 445 & \best{2.29}
& & 15.03 \\
Ackerman 3 9 & 4093 & 8G
  & 116.946 & 1055 & 90
  & 73.624 & 883 & 172 & 1.59
  & 59.421 & 756 & 248 & 1.97
  & 52.597 & 658 & 322 & 2.22
  & 49.011 & 579 & 392 & 2.39
  & 47.341 & 513 & 458 & \best{2.47}
& & 16.3 \\
\hline
Ackerman 3 10 & 8189 & 8G
  & 501.817 & 971 & 91
  & 326.369 & 774 & 175 & 1.54
  & 254.119 & 686 & 253 & 1.97
  & 222.919 & 606 & 327 & 2.25
  & 201.495 & 550 & 398 & 2.49
  & 191.309 & 500 & 461 & \best{2.62}
  & 28.677 & 17.5 \\
\hline
\hline
Ack 3 6 & 509 & \defaultHeap
  & 1.161 & 2201 & 42
  & 0.686 & 1456 & 109 & 1.70
  & 0.526 & 1189 & 174 & 2.21
  & 0.475 & 934 & 245 & 2.44
  & 0.439 & 780 & 317 & \best{2.64}
  & 0.443 & 642 & 383 & 2.62
& 0.078 & 14.9 \\
Ack 3 6 & 509 & 20M
  & 0.710 & 1691 & 90
  & 0.524 & 1196 & 174 & 1.35
  & 0.463 & 926 & 254  & 1.53
  & 0.433 & 774 & 325 & 1.64
  & \best{0.428} & 645 & 394 & \best{1.66}
  & 0.502 & 575 & 377 & 1.41
& & 9.10 \\
\hline
Ack 3 7 & 1021 & \defaultHeap
  & 6.412 & 2117 & 32
  & 3.552 & 1427 & 85 & 1.81
  & 2.473 & 1224 & 143 & 2.59
  & 2.187 & 956 & 207  & 2.93
  & 1.933 & 815 & 275 & 3.32
  & 1.835 & 686 & 344 &  \best{3.49}
& 0.310 & 20.7  \\
Ack 3 7 & 1021 & 40M
  & 3.023 & 1612 & 89
  & 2.180 & 1157 & 172 & 1.39
  & 1.909 & 908 & 250 & 1.58
  & 1.709 & 782 & 325 & 1.77
  & \best{1.698} & 646 & 395 & \best{1.78}
  & 1.896 & 590 & 388 & 1.59
& & 9.75 \\
\hline
Ack 3 8 & 2045 & \defaultHeap
  & 31.762 & 2039 & 27
  & 18.775 & 1357 & 68 & 1.69
  & 12.488 & 1206 & 115 & 2.54
  & 10.555 & 954 & 172 & 3.01
  & 9.132 & 810 & 235 & 3.48
  & 8.543 & 695 & 292 & \best{3.72}
& 1.109 & 28.64 \\ %
Ack 3 8 & 2045 & 60M
  & 13.675 & 1445 & 88
  & 9.743 & 1053 & 169 & 1.40
  & 8.044 & 875 & 247 & 1.70
  & 7.291 & 742 & 321 & 1.88
  & \best{6.914} & 637 & 394 & \best{1.98}
  & 7.027 & 529 & 467 & 1.95
& & 12.3 \\
\hline
\hline
fib 20 &   6765 & 1G
  & 0.587 & 714 & 94 
  & 0.390 & 602 & 169 & 1.50 
  & 0.325 & 527 & 234 & 1.80
  & 0.291 & 470 & 293 & 2.02
  & 0.266 & 428 & 353 & 2.21
  & 0.260 & 401 & 385 & \best{2.26}
  & 0.030 & 19.6 \\
\hline
fib 25 &  75025 & 1G
  & 8.832 & 838 & 56
  & 5.531 & 674 & 112 & 1.60
  & 4.250 & 576 & 171 & 2.08
  & 3.754 & 515 & 216 & 2.35
  & 3.373 & 456 & 272 & 2.62
  & 3.141 & 413 & 323 & \best{2.81}
  & 0.513 & 17.2 \\
fib 25 & 75025 & \bf {2G}
  & 7.096 & 797 & 74
  & 4.603 & 637 & 144 & 1.54
  & 3.685 & 562 & 205 & 1.93
  & 3.398 & 473 & 264 & 2.09
  & 3.082 & 446 & 310 & 2.30
  & \best{2.958} & 402 & 359 & \best{2.40}
  & & 13.8 \\
fib 25 & 75025 & 4G
  & 7.321 & 677 & 85
  & 4.709 & 568 & 160 & 1.55
  & 3.910 & 486 & 228 & 1.87
  & 3.445 & 432 & 291 & 2.13
  & 3.226 & 404 & 333 & 2.27
  & \best{2.958} & 365 & 406 & \best{2.47}
  & & 14.3  \\
fib 25 & 75025 & 8G
  & 7.538 & 667 & 86
  & 4.907 & 563 & 157 & 1.54
  & 4.026 & 501 & 217 & 1.87
  & 3.667 & 442 & 271 & 2.06
  & 3.318 & 401 & 331 & \best{2.27}
  & 3.329 & 359 & 372 & 2.26
  & & 14.7 \\
fib 25 & 75025 & 12G
  & 7.589 & 684 & 84
  & 5.023 & 580 & 151 & 1.51
  & 4.218 & 502 & 209 & 1.80
  & 3.804 & 443 & 264 & 2.00
  & 3.428 & 409 & 318 & 2.21
  & 3.226 & 382 & 362 & \best{2.35}
& & 14.8 \\
\hline
fib 28 & 317811 & 2G
  & 89.117 & 834 & 25
  & 54.464 & 682 & 51 & 1.64
  & 54.409 & 532 & 63 & 1.64
  & 48.101 & 493 & 79 & 1.85
  & 35.036 & 452 & 115 & \best{2.54}
  & 44.302 & 406 & 101 & 2.011
  & \multicolumn{2}{c|}{SegFault}
  \\
fib 28 & 317811 & 4G
  & 42.922 & 826 & 52
  & 25.890 & 649 & 109 & 1.66
  & 21.034 & 567 & 141 & 2.04
  & 19.416 & 501 & 189 & 2.21
  & 16.953 & 425 & 256 & 2.53
  & 16.310 & 411 & 283 & \best{2.63}
  & \multicolumn{2}{c|}{{}}
  \\
fib 28 & 317811 & 8G
  & 35.092 & 777 & 68
  & 22.117 & 627 & 133 & 1.59
  & 18.213 & 541 & 189 & 1.93
  & 16.366 & 482 & 236 & 2.144
  & 14.749 & 444 & 290 & 2.38
  & 14.093 & 397 & 335 & \best{2.49}
  & \multicolumn{2}{c|}{{}}
  \\
fib 28 & 317811 & \bf 12G
  & 34.283 & 750 & 72
  & 22.027 & 613 & 138 & 1.56
  & 17.603 & 551 & 193 & 1.95
  & 16.007 & 465 & 253 & 2.14
  & 14.546 & 430 & 301 & 2.36
  & \best{13.759} & 386 & 355 & \best{2.49}
  & \multicolumn{2}{c|}{{}}
  \\
\hline
 fib 30 & 832040 & 12G
  & 116.444 & 804 & 53
  & 76.595 & 624 & 104 & 1.52
  & 62.474 & 559 & 142 & 1.86
  & 55.308 & 495 & 182 & 2.11
  & 49.664 & 454 & 222 & 2.34
  & 45.192 & 406 & 273 & \best{2.58}
  & \multicolumn{2}{c|}{SegFault}
  \\
\hline
\hline
fibonacci 20 & 6765 & 1G
  & 0.269 & 759 & 91
  & 0.185 & 619 & 163 & 1.45
  & 0.154 & 551 & 223 & 1.75
  & 0.141 & 486 & 275 & 1.91
  & 0.130 & 456 & 320 & 2.07
  & 0.125 & 417 & 366 & \best{2.15}
  & 0.018 & 14.9   \\
fibonacci 25 & 75025 & 2G
  & 2.740 & 742 & 95
  & 1.841 & 590 & 178 & 1.49
  & 1.459 & 526 & 255 & 1.88
  & 1.324 & 451 & 327 & 2.07
  & 1.172 & 404 & 400 & 2.34
  & 1.134 & 377 & 461 & \best{2.42}
  & 0.172 & 15.9 \\
fibonacci 28 & 317811 & 8G
  & 11.752 & 724 & 90
  & 7.926 & 583 & 167 & 1.48
  & 5.928 & 508 & 258 & 1.98
  & 5.460 & 446 & 321 & 2.15
  & 4.934 & 404 & 395 & 2.38
  & 4.519 & 374 & 469 & \best{2.60}
  & 0.721 & 16.3 \\
fibonacci 30 & 832040 & 8G
  & 47.947 & 1048 & 45
  & 30.137 & 707 & 96 & 1.59
  & 23.071 & 608 & 144 & 2.08
  & 17.589 & 513 & 222 & 2.73
  & 15.243 & 464 & 282 & \best{3.15}
  & 16.191 & 422 & 292 & 2.96
  &  1.858 & 25.8 \\
\hline
\hline
sort200 &  & \defaultHeap
  & 0.123 & 2226 & 46
  & 0.092 & 1443 & 94 & 1.34
  & 0.085 & 1171 & 126 & 1.45
  & 0.080 & 925 & 168 & \best{1.54}
  & 0.080 & 766 & 204 & \best{1.54}
  & 0.080 & 662 & 236 & \best{1.54}
  & 0.012 & 10.3 \\
sort300 &  & \defaultHeap
  & 0.255 & 2114 & 44
  & 0.191 & 1341 & 93 & 1.34
  & 0.155 & 1115 & 138 & 1.65
  & 0.152 & 874 & 180 & 1.68
  & 0.159 & 695 & 217 & 1.60
  & 0.145 & 626 & 264 & \best{1.76}
  & 0.023 & 11.1 \\
sort400 &  & \defaultHeap
  & 0.466 & 2039 & 41
  & 0.328 & 1291 & 92 & 1.42
  & 0.254 & 1084 & 142 & 1.83
  & 0.243 & 851 & 190 & 1.92
  & 0.223 & 741 & 237 & 2.09
  & 0.218 & 622 & 289 & \best{2.13}
  & 0.037 & 12.6 \\
sort500 &  & \defaultHeap
  & 0.802 & 1924 & 38 
  & 0.543 & 1233 & 87 & 1.48
  & 0.400 & 1057 & 137 & 2.01
  & 0.376 & 826 & 187 & 2.13
  & 0.345 & 709 & 238 & 2.32
  & 0.332 & 619 & 283 & \best{2.41}
  & 0.060 & 13.4 \\
sortC600 &  & 50M
  & 0.570 & 1504 & 91
  & 0.439 & 1054 & 169 & 1.30
  & 0.367 &  875 & 243 & 1.55
  & 0.349 & 714 & 313 & 1.63
  & 0.335 & 607 & 384 & 1.70
  & 0.326 & 534 & 449 & \best{1.75}
  & 0.071 & 8.03 \\
sortC700 &  & 50M
  & 0.765 & 1495 & 91
  & 0.586 & 1044 & 170 & 1.31
  & 0.492 & 862 & 246 & 1.55
  & 0.460 & 708 & 319 & 1.66
  & 0.430 & 627 & 387 & 1.78
  & 0.429 & 530 & 458 & \best{1.78}
  & 0.095 & 8.05\\
sortC800 &  & 50M
  & 1.002 & 1474 & 91
  & 0.756 & 1039 & 170 & 1.31
  & 0.625 & 869 & 246 & 1.55
  & 0.586 & 714 & 320 & 1.66
  & 0.558 & 613 & 391 & 1.78
  & 0.557 & 526 & 458 & \best{1.78}
  & 0.121 & 8.05 \\
sortC900 &  & 50M
  & 1.274 & 1451 & 90
  & 0.955 & 1032 & 170 & 1.33
  & 0.798 & 850 & 248 & 1.60
  & 0.741 & 707 & 320 & 1.72
  & 0.698 & 620 & 393 & 1.83
  & 0.688 & 530 & 460 & \best{1.85}
  & 0.155 & 8.22 \\
\hline
sortC1000 &  & \defaultHeap
  & 3.810 & 1741 & 31
  & 2.307 & 1158 & 77 & 1.65
  & 1.602 & 995 & 128 & 2.38
  & 1.404 & 790 & 184 & 2.71
  & 1.215 & 682 & 247 & 3.14
  & 1.105 & 600 & 309 & \best{3.45}
  & 0.196 & 19.4 \\
\hline
sortC1000 &  & 100M
  & 1.599 & 1410 & 91
  & 1.196 & 997 & 173 & 1.34
  & 0.999 & 827 & 249 & 1.60
  & 0.902 & 700 & 326 & 1.77
  & 0.853 & 605 & 399 & 1.87
  & 0.845 & 520 & 469 & \best{1.89}
  & 0.196 & 8.16 \\
sortC2000 &  & 100M
  & 6.805 & 1292 & 90
  & 4.915 & 936 & 172 & 1.38
  & 3.939 & 800 & 251 & 1.73
  & 3.599 & 671 & 328 & 1.89
  & 3.323 & 592 & 404 & 2.05
  & 3.239 & 510 & 480 & \best{2.10}
  & \multicolumn{2}{c|}{\scalebox{0.7}[1.0]{StackOverflow}} \\
sortC3000 &  & 100M
  & 16.932 & 1174 & 88
  & 11.656 & 892 & 169 & 1.45
  & 9.261 & 768 & 248 & 1.83
  & 8.344 & 654 & 323 & 2.03
  & 7.674 & 579 & 397 & 2.21
  & 7.430 & 503 & 472 & \best{2.28}
  & \multicolumn{2}{c|}{\scalebox{0.7}[1.0]{StackOverflow}} \\
sortC4000 &  & 100M
  & 32.337 & 1108 & 87
  & 21.991 & 854 & 166 & 1.47
  & 17.454 & 737 & 242 & 1.85
  & 15.580 & 634 & 315 & 2.08
  & 14.112 & 567 & 390 & 2.29
  & 13.638 & 497 & 460 & \best{2.37}
  & \multicolumn{2}{c|}{\scalebox{0.7}[1.0]{StackOverflow}} \\
sortC5000 &  & 100M
  & 54.673 & 1037 & 85
  & 36.384 & 830 & 161 & 1.50
  & 28.286 & 722 & 238 & 1.93
  & 25.387 & 620 & 308 & 2.15
  & 22.902 & 558 & 380 & 2.38
  & 21.701 & 492 & 455 & \best{2.52}
  & \multicolumn{2}{c|}{\scalebox{0.7}[1.0]{StackOverflow}} \\
sortC10000 &  & 1G
  & 247.709 & 839 & 93
  & 165.529 & 650 & 180 & 1.50
  & 131.459 & 554 & 265 & 1.88
  & 110.329 & 506 & 346 & 2.25
  & 97.315 & 472 & 420 & 2.55
  & 89.930s & 424 & 506 & \best{2.75}
  & \multicolumn{2}{c|}{\scalebox{0.7}[1.0]{StackOverflow}} \\
\hline
\end{tabular}
}}
\caption{Benchmarks for \Inets{} programs}\label{inetTable}
\end{table}

Just adding cores to a \ensuremath{\VarId{RunInets}} run without any heap settings
(see the ``\defaultHeap{}'' rows)
appears to produce relatively nice speed-ups for fine-grained
parallelism, but one has to be aware that the single-core execution in
that case typically was spending a far larger portion of its time in garbage
collection than the multi-core versions.
(This applies also for ``relatively small'' fixed heaps.)

It appears to be more honest to consider the speed-ups
compared to single-core executions with a ``good'' fixed heap setting;
the fastest runs on our six-core machine with our parallel interpreter
all use five or six cores,
and tend to take only about five to six times as long as the compiled \Inets{}
runs on a single core.

{\small
(For reasons I have not investigated, the \Inets{}-compiled
executables
crashed for the larger \textsf{fib.inet} runs after producing partial
output;
on a modified version (\textsf{fibNat.inet}) that converts results from
unary representation to \ensuremath{\VarId{int}} attributes, all \Inets{} runs crashed. 
For \textsf{sort.inet}, the \Inets{} version was originally changed only by adding
longer argument lists to the start net; beyond 500 elements, this lead
to stack overflow errors in the \textsf{javacc}-generated parser.
Changing the start net definition to a sequence of equations
each adding a smaller chunk to the list allowed us to make some
progress, but beyond 1000 elements, a different stack overflow occurred.)
}

\section{Conclusion}

Interaction nets as an ``inherently parallel'' execution model
promise large speed-ups via parallelisation,
but accessible platforms for experimentation are still missing.

Using Concurrent Haskell to implement interaction nets
understood as an execution mechanism,
we achieved a simple and easily understandable implementation,
the entire core of which could be presented in just a bit more than three pages of literate
code.
By having added support for the \Inets{} file format,
we enable experimentation with interaction net definitions
in the shape used by most of the current interaction net literature
--- with the restriction that a consistent polarity assignment must be
possible (which is also one of the conditions of Lafont
\cite{Lafont-1990} for deadlock safety).

Keeping in mind that, in our straight-forward ultrafine-grained implementation,
the concurrent interaction net rules reduce a heavily shared structure,
and given that we made no effort to enable coarse-grain parallelism,
the speed-ups achieved on the usual microbenchmarks are actually
surprisingly good, and we expect even better behaviour on rules with
larger right-hand sides that give rise to more sparsely connected nets.

{
\bibliographystyle{eptcsalpha} 
\bibliography{strings,parnas,ref,crossrefs}

\newcommand{\etalchar}[1]{$^{#1}$}
\begin{thebibliography}{AFK{\etalchar{+}}11}
\providecommand{\bibitemdeclare}[2]{}
\providecommand{\surnamestart}{}
\providecommand{\surnameend}{}
\providecommand{\urlprefix}{Available at }
\providecommand{\url}[1]{\texttt{#1}}
\providecommand{\href}[2]{\texttt{#2}}
\providecommand{\urlalt}[2]{\href{#1}{#2}}
\providecommand{\doi}[1]{doi:\urlalt{http://dx.doi.org/#1}{#1}}
\providecommand{\bibinfo}[2]{#2}

\bibitemdeclare{inproceedings}{Andrei-Fernandez-Kirchner-Melancon-Namet-Pinaud-2011}
\bibitem[AFK{\etalchar{+}}11]{Andrei-Fernandez-Kirchner-Melancon-Namet-Pinaud-2011}
\bibinfo{author}{Oana \surnamestart Andrei\surnameend},
  \bibinfo{author}{Maribel \surnamestart Fern{\'{a}}ndez\surnameend},
  \bibinfo{author}{H{\'e}l{\`e}ne \surnamestart Kirchner\surnameend},
  \bibinfo{author}{Guy \surnamestart Melan{\c{c}}on\surnameend},
  \bibinfo{author}{Olivier \surnamestart Namet\surnameend} \&
  \bibinfo{author}{Bruno \surnamestart Pinaud\surnameend}
  (\bibinfo{year}{2011}): \emph{\bibinfo{title}{{PORGY}: Strategy-driven
  interactive transformation of graphs}}.
\newblock In \bibinfo{editor}{Rachid \surnamestart Echahed\surnameend}, editor:
  {\sl \bibinfo{booktitle}{{TERMGRAPH 2011}, International Workshop on Term
  Graph Rewriting}}, {\sl \bibinfo{series}{Electronic Proceedings in Computer
  Science}}~\bibinfo{volume}{48}, pp. \bibinfo{pages}{54--68},
  \doi{10.4204/EPTCS.48.7}.

\bibitemdeclare{incollection}{Anand-Kahl-2009a}
\bibitem[AK09]{Anand-Kahl-2009a}
\bibinfo{author}{Christopher~K. \surnamestart Anand\surnameend} \&
  \bibinfo{author}{Wolfram \surnamestart Kahl\surnameend}
  (\bibinfo{year}{2009}): \emph{\bibinfo{title}{Synthesizing and Verifying
  Multicore Parallelism in Categories of Nested Code Graphs}}.
\newblock In \bibinfo{editor}{Michael \surnamestart Alexander\surnameend} \&
  \bibinfo{editor}{William \surnamestart Gardner\surnameend}, editors: {\sl
  \bibinfo{booktitle}{Process Algebra for Parallel and Distributed
  Processing}}, chapter~\bibinfo{chapter}{1}, {\sl \bibinfo{series}{CRC
  Computational Science Series}}~\bibinfo{volume}{2},
  \bibinfo{publisher}{Chapman \& Hall}, pp. \bibinfo{pages}{3--45},
  \doi{10.1201/9781420064872.pt1}.

\bibitemdeclare{article}{Almeida-Pinto-Vilaca-2008}
\bibitem[APV08]{Almeida-Pinto-Vilaca-2008}
\bibinfo{author}{Jos{\'e}~Bacelar \surnamestart Almeida\surnameend},
  \bibinfo{author}{Jorge~Sousa \surnamestart Pinto\surnameend} \&
  \bibinfo{author}{Miguel \surnamestart Vila{\c{c}}a\surnameend}
  (\bibinfo{year}{2008}): \emph{\bibinfo{title}{A Tool for Programming with
  Interaction Nets}}.
\newblock {\sl \bibinfo{journal}{ENTCS}} \bibinfo{volume}{219}, pp.
  \bibinfo{pages}{83--96}, \doi{10.1016/j.entcs.2008.10.036}.
\newblock \bibinfo{note}{Proc.~Eighth International Workshop on Rule Based
  Programming (RULE 2007)}.

\bibitemdeclare{article}{Banach-Papadopoulos-1997}
\bibitem[BP97]{Banach-Papadopoulos-1997}
\bibinfo{author}{Richard \surnamestart Banach\surnameend} \&
  \bibinfo{author}{George~A. \surnamestart Papadopoulos\surnameend}
  (\bibinfo{year}{1997}): \emph{\bibinfo{title}{A Study of Two Graph Rewriting
  Formalisms: Interaction Nets and {MONSTR}}}.
\newblock {\sl \bibinfo{journal}{Journal of Programming Languages}}
  \bibinfo{volume}{5}, pp. \bibinfo{pages}{210--231}.

\bibitemdeclare{article}{CirsteaH-Faure-Fernandez-Mackie-Sinot-2007}
\bibitem[CFF{\etalchar{+}}07]{CirsteaH-Faure-Fernandez-Mackie-Sinot-2007}
\bibinfo{author}{Horatiu \surnamestart Cirstea\surnameend},
  \bibinfo{author}{Germain \surnamestart Faure\surnameend},
  \bibinfo{author}{Maribel \surnamestart Fern{\'a}ndez\surnameend},
  \bibinfo{author}{Ian \surnamestart Mackie\surnameend} \&
  \bibinfo{author}{Fran{\c{c}}ois-R{\'e}gis \surnamestart Sinot\surnameend}
  (\bibinfo{year}{2007}): \emph{\bibinfo{title}{From Functional Programs to
  Interaction Nets via the Rewriting Calculus}}.
\newblock {\sl \bibinfo{journal}{ENTCS}}
  \bibinfo{volume}{174}(\bibinfo{number}{10}), pp. \bibinfo{pages}{39--56},
  \doi{10.1016/j.entcs.2007.02.046}.
\newblock \bibinfo{note}{Proceedings of the Sixth International Workshop on
  Reduction Strategies in Rewriting and Programming (WRS 2006)}.

\bibitemdeclare{inproceedings}{Erkoek-Launchbury-2002}
\bibitem[EL02]{Erkoek-Launchbury-2002}
\bibinfo{author}{Levent \surnamestart Erk{\"o}k\surnameend} \&
  \bibinfo{author}{John \surnamestart Launchbury\surnameend}
  (\bibinfo{year}{2002}): \emph{\bibinfo{title}{A recursive do for {Haskell}}}.
\newblock In \bibinfo{editor}{Manuel \surnamestart Chakravarty\surnameend},
  editor: {\sl \bibinfo{booktitle}{Proc.\null{} Haskell Workshop 2002}},
  \bibinfo{publisher}{ACM Press}, pp. \bibinfo{pages}{29--37},
  \doi{10.1145/581690.581693}.

\bibitemdeclare{misc}{deFalco-2006}
\bibitem[Fal06]{deFalco-2006}
\bibinfo{author}{Marc \surnamestart de~Falco\surnameend}
  (\bibinfo{year}{2006}): \emph{\bibinfo{title}{Interaction Nets Laboratory}}.
\newblock \urlprefix\url{http://inl.sourceforge.net/}.

\bibitemdeclare{misc}{INets}
\bibitem[HJ12]{INets}
\bibinfo{author}{Abubakar \surnamestart Hassan\surnameend} \&
  \bibinfo{author}{Eugen \surnamestart Jiresch\surnameend}
  (\bibinfo{year}{2012}): \emph{\bibinfo{title}{Interaction Nets Programming
  Language}}.
\newblock \bibinfo{howpublished}{\url{https://gna.org/projects/inets/},
  \url{https://gna.org/svn/?group=inets}, last accessed 2015-02-06}.
\newblock \bibinfo{note}{(Source code for the ``Inets'' system.)}.

\bibitemdeclare{inproceedings}{Hassan-Jiresch-SatoShinya-2010}
\bibitem[HJS09]{Hassan-Jiresch-SatoShinya-2010}
\bibinfo{author}{Abubakar \surnamestart Hassan\surnameend},
  \bibinfo{author}{Eugen \surnamestart Jiresch\surnameend} \&
  \bibinfo{author}{Shinya \surnamestart Sato\surnameend}
  (\bibinfo{year}{2009}): \emph{\bibinfo{title}{An Implementation of Nested
  Pattern Matching in Interaction Nets}}.
\newblock In \bibinfo{editor}{Ian \surnamestart Mackie\surnameend} \&
  \bibinfo{editor}{Anamaria~Martins \surnamestart Moreira\surnameend}, editors:
  {\sl \bibinfo{booktitle}{Proceedings Tenth International Workshop on
  Rule-Based Programming, {RULE} 2009, Bras{\'{\i}}lia, Brazil, 28th June
  2009.}}, {\sl \bibinfo{series}{{EPTCS}}}~\bibinfo{volume}{21}, pp.
  \bibinfo{pages}{13--25}, \doi{10.4204/EPTCS.21.2}.

\bibitemdeclare{article}{Hassan-Mackie-Sato-2009}
\bibitem[HMS09]{Hassan-Mackie-Sato-2009}
\bibinfo{author}{Abubakar \surnamestart Hassan\surnameend},
  \bibinfo{author}{Ian \surnamestart Mackie\surnameend} \&
  \bibinfo{author}{Shinya \surnamestart Sato\surnameend}
  (\bibinfo{year}{2009}): \emph{\bibinfo{title}{Compilation of Interaction
  Nets}}.
\newblock {\sl \bibinfo{journal}{ENTCS}}
  \bibinfo{volume}{253}(\bibinfo{number}{4}), pp. \bibinfo{pages}{73--90},
  \doi{10.1016/j.entcs.2009.10.018}.
\newblock \bibinfo{note}{Proc.~TERMGRAPH 2009}.

\bibitemdeclare{inproceedings}{Hassan-Mackie-Sato-2010}
\bibitem[HMS10]{Hassan-Mackie-Sato-2010}
\bibinfo{author}{Abubakar \surnamestart Hassan\surnameend},
  \bibinfo{author}{Ian \surnamestart Mackie\surnameend} \&
  \bibinfo{author}{Shinya \surnamestart Sato\surnameend}
  (\bibinfo{year}{2010}): \emph{\bibinfo{title}{A lightweight abstract machine
  for interaction nets}}.
\newblock In \bibinfo{editor}{Jochen \surnamestart K\"uster\surnameend} \&
  \bibinfo{editor}{Emilio \surnamestart Tuosto\surnameend}, editors: {\sl
  \bibinfo{booktitle}{Proc.~GT-VMT 2010}}, {\sl
  \bibinfo{series}{ECEASST}}~\bibinfo{volume}{29}, pp.
  \bibinfo{pages}{9.1--9.12}.
\newblock
  \urlprefix\url{http://journal.ub.tu-berlin.de/eceasst/article/view/416}.

\bibitemdeclare{article}{Hoffmann-Plump-1991}
\bibitem[HP91]{Hoffmann-Plump-1991}
\bibinfo{author}{Berthold \surnamestart Hoffmann\surnameend} \&
  \bibinfo{author}{Detlef \surnamestart Plump\surnameend}
  (\bibinfo{year}{1991}): \emph{\bibinfo{title}{Implementing Term Rewriting by
  Jungle Evaluation}}.
\newblock {\sl \bibinfo{journal}{Informatique th\'eorique et
  applications/Theoretical Informatics and Applications}}
  \bibinfo{volume}{25}(\bibinfo{number}{5}), pp. \bibinfo{pages}{445--472}.

\bibitemdeclare{inproceedings}{Jiresch-2014}
\bibitem[Jir14]{Jiresch-2014}
\bibinfo{author}{Eugen \surnamestart Jiresch\surnameend}
  (\bibinfo{year}{2014}): \emph{\bibinfo{title}{Towards a GPU-based
  Implementation of Interaction Nets}}.
\newblock In \bibinfo{editor}{Benedikt \surnamestart L{\"o}we\surnameend} \&
  \bibinfo{editor}{Glynn \surnamestart Winskel\surnameend}, editors: {\sl
  \bibinfo{booktitle}{8th International Workshop on Developments in
  Computational Models, {DCM 2012}}}, {\sl \bibinfo{series}{EPTCS}}
  \bibinfo{volume}{143}, pp. \bibinfo{pages}{41--53},
  \doi{10.4204/EPTCS.143.4}.

\bibitemdeclare{inproceedings}{Kahl-Anand-Carette-2005}
\bibitem[KAC06]{Kahl-Anand-Carette-2005}
\bibinfo{author}{Wolfram \surnamestart Kahl\surnameend},
  \bibinfo{author}{Christopher~Kumar \surnamestart Anand\surnameend} \&
  \bibinfo{author}{Jacques \surnamestart Carette\surnameend}
  (\bibinfo{year}{2006}): \emph{\bibinfo{title}{Control-Flow Semantics for
  Assembly-Level Data-Flow Graphs}}.
\newblock In \bibinfo{editor}{Wendy \surnamestart McCaull\surnameend},
  \bibinfo{editor}{Michael \surnamestart Winter\surnameend} \&
  \bibinfo{editor}{Ivo \surnamestart D{\"u}ntsch\surnameend}, editors: {\sl
  \bibinfo{booktitle}{8th Intl.\null{} Seminar on Relational Methods in
  Computer Science, {RelMiCS 8, Feb.\null{} 2005}}}, {\sl
  \bibinfo{series}{LNCS}} \bibinfo{volume}{3929},
  \bibinfo{publisher}{Springer}, pp. \bibinfo{pages}{147--160},
  \doi{10.1007/11734673\_12}.

\bibitemdeclare{incollection}{Kennaway-Klop-Sleep-deVries-1993a}
\bibitem[KKSV93]{Kennaway-Klop-Sleep-deVries-1993a}
\bibinfo{author}{J.R. \surnamestart Kennaway\surnameend}, \bibinfo{author}{J.W.
  \surnamestart Klop\surnameend}, \bibinfo{author}{M.R. \surnamestart
  Sleep\surnameend} \& \bibinfo{author}{F.J. \surnamestart de~Vries\surnameend}
  (\bibinfo{year}{1993}): \emph{\bibinfo{title}{An Introduction to Term Graph
  Rewriting}}.
\newblock In \bibinfo{editor}{M.R. \surnamestart Sleep\surnameend},
  \bibinfo{editor}{M.J. \surnamestart Plasmeijer\surnameend} \&
  \bibinfo{editor}{M.C.J.D. \surnamestart van Eekelen\surnameend}, editors:
  {\sl \bibinfo{booktitle}{Term Graph Rewriting: Theory and Practice}},
  chapter~\bibinfo{chapter}{1}, \bibinfo{publisher}{Wiley}, pp.
  \bibinfo{pages}{1--14}.

\bibitemdeclare{inproceedings}{Lafont-1990}
\bibitem[Laf90]{Lafont-1990}
\bibinfo{author}{Yves \surnamestart Lafont\surnameend} (\bibinfo{year}{1990}):
  \emph{\bibinfo{title}{Interaction Nets}}.
\newblock In: {\sl \bibinfo{booktitle}{17th POPL}}, \bibinfo{publisher}{ACM},
  \bibinfo{address}{New York, NY, USA}, pp. \bibinfo{pages}{95--108},
  \doi{10.1145/96709.96718}.

\bibitemdeclare{incollection}{Lippi-2002}
\bibitem[Lip02]{Lippi-2002}
\bibinfo{author}{Sylvain \surnamestart Lippi\surnameend}
  (\bibinfo{year}{2002}): \emph{\bibinfo{title}{in$^2$: A Graphical Interpreter
  for Interaction Nets}}.
\newblock In \bibinfo{editor}{Sophie \surnamestart Tison\surnameend}, editor:
  {\sl \bibinfo{booktitle}{{RTA 2002}}}, {\sl \bibinfo{series}{LNCS}}
  \bibinfo{volume}{2378}, \bibinfo{publisher}{Springer},
  \bibinfo{address}{Berlin Heidelberg}, pp. \bibinfo{pages}{380--385},
  \doi{10.1007/3-540-45610-4\_29}.

\bibitemdeclare{inproceedings}{Mackie-1998}
\bibitem[Mac98]{Mackie-1998}
\bibinfo{author}{Ian \surnamestart Mackie\surnameend} (\bibinfo{year}{1998}):
  \emph{\bibinfo{title}{{YALE}: Yet Another Lambda Evaluator Based on
  Interaction Nets}}.
\newblock In: {\sl \bibinfo{booktitle}{Proceedings of the Third ACM SIGPLAN
  International Conference on Functional Programming}}, \bibinfo{series}{ICFP
  '98}, \bibinfo{publisher}{ACM}, \bibinfo{address}{New York, NY, USA}, pp.
  \bibinfo{pages}{117--128}, \doi{10.1145/289423.289434}.

\bibitemdeclare{article}{Mackie-2005}
\bibitem[Mac05]{Mackie-2005}
\bibinfo{author}{Ian \surnamestart Mackie\surnameend} (\bibinfo{year}{2005}):
  \emph{\bibinfo{title}{Towards a Programming Language for Interaction Nets}}.
\newblock {\sl \bibinfo{journal}{ENTCS}}
  \bibinfo{volume}{127}(\bibinfo{number}{5}), pp. \bibinfo{pages}{133--151},
  \doi{10.1016/j.entcs.2005.02.015}.
\newblock \bibinfo{note}{Proc.~TERMGRAPH 2004}.

\bibitemdeclare{incollection}{Pinto-2001RTA}
\bibitem[Pin01]{Pinto-2001RTA}
\bibinfo{author}{Jorge~Sousa \surnamestart Pinto\surnameend}
  (\bibinfo{year}{2001}): \emph{\bibinfo{title}{Parallel Evaluation of
  Interaction Nets with {MPINE}}}.
\newblock In \bibinfo{editor}{Aart \surnamestart Middeldorp\surnameend},
  editor: {\sl \bibinfo{booktitle}{Rewriting Techniques and Applications, {RTA
  2001}}}, {\sl \bibinfo{series}{LNCS}} \bibinfo{volume}{2051},
  \bibinfo{publisher}{Springer}, pp. \bibinfo{pages}{353--356},
  \doi{10.1007/3-540-45127-7\_26}.

\bibitemdeclare{inproceedings}{PeytonJones-Gordon-Finne-1996}
\bibitem[PJGF96]{PeytonJones-Gordon-Finne-1996}
\bibinfo{author}{Simon~L. \surnamestart Peyton~Jones\surnameend},
  \bibinfo{author}{Andrew \surnamestart Gordon\surnameend} \&
  \bibinfo{author}{Sigbjorn \surnamestart Finne\surnameend}
  (\bibinfo{year}{1996}): \emph{\bibinfo{title}{Concurrent Haskell}}.
\newblock In: {\sl \bibinfo{booktitle}{23rd POPL}}, \bibinfo{publisher}{acm
  press}, pp. \bibinfo{pages}{295--308}, \doi{10.1145/237721.237794}.

\bibitemdeclare{article}{Pedicini-Quaglia-2007}
\bibitem[PQ07]{Pedicini-Quaglia-2007}
\bibinfo{author}{Marco \surnamestart Pedicini\surnameend} \&
  \bibinfo{author}{Francesco \surnamestart Quaglia\surnameend}
  (\bibinfo{year}{2007}): \emph{\bibinfo{title}{PELCR: Parallel Environment for
  Optimal Lambda-calculus Reduction}}.
\newblock {\sl \bibinfo{journal}{ACM Trans.\null{} Computational Logic}}
  \bibinfo{volume}{8}(\bibinfo{number}{3}), \doi{10.1145/1243996.1243997}.

\end{thebibliography}
}

\end{document}